\documentclass[leqno, 11pt]{article}
\usepackage{times}
\usepackage{amsmath}
\usepackage{amssymb}
\usepackage{setspace}

\numberwithin{equation}{section}

\newtheorem{theorem}{Theorem}[section]
\newtheorem{definition}[theorem]{Definition}

\newtheorem{lemma}[theorem]{Lemma}
\newtheorem{corollary}[theorem]{Corollary}

\begin{document}

\date{}

\title{Necessity as justified truth}

\author{Steffen Lewitzka\thanks{Universidade Federal da Bahia -- UFBA,
Instituto de Matem\'atica,
Departamento de Ci\^encia da Computa\c c\~ao,
Campus de Ondina,
40170-110 Salvador -- BA,
Brazil,
steffen@dcc.ufba.br}}

\maketitle

\begin{abstract}
We present a logic for the reasoning about necessity and justifications which is independent from relational semantics. 
We choose the concept of \textit{justification} -- coming from a class of \textit{Justification Logics} (Artemov 2008, Fitting 2009) -- as the primitive notion on which the concept of necessity is based. Our axiomatization extends Suszko's non-Fregean logic SCI (Brown, Suszko 1972) by basic axioms from \textit{Justification Logic}, axioms for quantification over propositions and over justifications, and some further principles. The core axiom is: $\varphi$ is necessarily true iff there is a justification for $\varphi$. That is, necessity is first-order definable by means of justifications. Instead of defining purely algebraic models in the style of (Brown, Suszko 1972) we extend the semantics investigated in (Lewitzka 2012) by some algebraic structure for dealing with justifications and prove soundness and completeness of our deductive system. Moreover, we are able to restore the modal logic principle of Necessitation if we add the axiom schema $\square\varphi\rightarrow\square\square\varphi$ and a rule of Axiom Necessitation to our system. As a main result, we show that the modal logics S4 and S5 can be captured by our semantics if we impose the corresponding modal logic principles as additional semantic constraints. This will follow from proof-theoretic considerations and from our completeness theorems. For the system S4 we present also a purely model-theoretic proof. 
\end{abstract}


\section{Introduction}

In an earlier paper \cite{lewsl} we presented an epistemic non-Fregean logic, called $\in_K$, which is independent of relational semantics and models knowledge and common knowledge avoiding all forms of the \textit{problem of logical omniscience}. Formulas are interpreted as elements of a given model-theoretic universe $M=TRUE\cup FALSE$ of true and false \textit{propositions}.\footnote{Propositions are in general given as abstract entities. A proposition $p$, as the element of a model-theoretic universe, can be identified with the set of all formulas that denote $p$.} Knowledge of an agent $i$ is explicitly given as a set of true propositions $TRUE_i\subseteq TRUE\subset M$. In general, the set $TRUE_i$ is not closed under rules of inference. Nevertheless, imposing specific constraints on the model-theoretic semantics, it can be closed under logical connectives, Modus Ponens (via a semantic counterpart of axiom $K$), positive and/or negative introspection, etc. That is, many standard epistemic principles can be restored and the reasoning capabilities of an agent may vary from completely ignorant to omniscient. The modal logic principle of Necessitation, however, is not investigated in \cite{lewsl}. A further question arising from \cite{lewsl} is that for an intuitive explanation or justification for knowledge. While (implicit) knowledge in modal epistemic logics derives in a rather intuitive way from the underlying possible worlds scenario ($\varphi$ is known iff $\varphi$ is true in all accessible worlds), knowledge in logic $\in_K$ is given \textit{a priori} as a set of propositions without any reasons provided by the semantics. We enrich here the approach developed in \cite{lewsl} by the intuitive (though abstract) concept of \textit{justification} coming from a class of Justification Logics which evolved from the Logic of Proofs introduced by Artemov \cite{art1, art2}. The notions of necessity/knowledge then are based on the primitive concept of justification. In fact, by introducing quantification over justifications we are able to define necessity/knowledge by means of a first-order existential formula. Justification logics are classical propositional logics augmented with justification assertions of the form $t:\varphi$ which read as ``$t$ is a justification for $\varphi$". In an epistemic framework, such an assertion reads as ``$\varphi$ is known for the explicit reason $t$". Under this interpretation, justification logics can be seen as logics of explicit knowledge addressing the problem of logical omniscience in a new way. Moreover, justification assertions turn epistemic logic more expressive and allow for a deeper epistemic analysis. For a detailed overview we refer the reader to \cite{art3, fit2, artfit}. The standard relational semantics for Justification Logic was developed by Fitting \cite{fit1}. In this paper, we apply only a few basic principles of classical Justification Logic and deviate from some others (for instance, we will work with a different rule of Axiom Necessitation). The basic principles that we adopt here are given by the following axiom schemas where $\cdot$ and $+$ are operations on justification terms: 
\begin{itemize}
\item$s:(\varphi\rightarrow\psi)\rightarrow (t:\varphi\rightarrow (s\cdot t):\psi)$ (application)
\item $s:\varphi\rightarrow (s+t):\varphi$ (weakening)
\item $t:\varphi\rightarrow (s+t):\varphi$ (weakening)
\item $(t:\varphi)\rightarrow\varphi$
\end{itemize}

We present a Hilbert-style deductive system (with Modus Ponens as the only rule) which contains the above given axioms and extends Suszko's basic non-Fregean logic SCI, \textit{the Sentential Calculus with Identity} \cite{blosus}, by quantification over propositions and quantification over justifications (or reasons), a total truth predicate, a connective for propositional reference, and a partial ordering on justifications. We define a semantics which extends the semantics studied in \cite{lewigpl} by some algebraic machinery for dealing with justifications. This style of non-Fregean semantics goes back to Str\"ater who designed $\in_T$-Logic \cite{str} as a first-order logic for the reasoning about propositional truth and propositional self-reference. $\in_T$-Logic does not involve inconsistencies despite of its total truth predicate and its capability to assert self-referential statements without restrictions. The Tarski biconditionals (Tarski's T-schema) hold and can be expressed in the object language (see, e.g., \cite{lewigpl, lewjlc} for a discussion). The axiomatization of $\in_T$-Logic given by Zeitz \cite{zei} can be seen as an extension of the axiomatization of Suszko's SCI.\footnote{Roughly speaking, $\in_T$-Logic = SCI + propositional quantifiers + a total truth predicate. The connection between $\in_T$-Logic and SCI remained unnoticed by Str\"ater and Zeitz. In fact, $\in_T$-Logic was developed independently from Suszko's non-Fregean logics proceeding from different assumptions and motivations.} Our Hilbert-style deductive system extends $\in_T$-Logic and therefore also SCI. On the other hand, $\in_T$-style semantics (which we will apply in this paper) differs essentially from the algebraic semantics given for SCI (see \cite{blosus}). The algebraic structure of an $\in_T$-model is not explicitly given by operations on a propositional universe but is imposed by the truth conditions and the structural properties of the semantic assignment function (called Gamma-function) of a model. In \cite{lewsl} it is shown that (in a quantifier-free setting) $\in_T$-style semantics is equivalent with the algebraic semantics in the style of Bloom/Suszko. An advantage of $\in_T$-style semantics is that models do not carry all the algebraic structure which is already implicitly given by the Gamma-function and the structure of the object language.

Our logic has all the expressive power of the $\in_T$-logic presented in \cite{lewigpl}. In particular, we may assert self-referential statements such as truth tellers, liars, contingent liars, etc. but also self-referential epistemic propositions in the object language. For instance, the equation $d_1\equiv\square d_1$ asserts that constant $d_1$ denotes the proposition ``I am necessarily true". Similarly, $d_2\equiv t: d_2$ asserts that $d_2$ denotes the proposition ``I am true for the explicit reason $t$". There exist models satisfying these equations. Equations such as $d_3\equiv (d_3:false)$ asserting a paradoxical statement can be formulated in the language but they are unsatisfiable. The fact that the liar proposition does not exist is expressed by the theorem $\neg\exists x.x\equiv (x:false)$.
The symbol $:$ in a formula $\varphi:true$ or $\varphi:false$ can be read as ``is element of". Thus, $\varphi:false$ reads as ``the proposition denoted by $\varphi$ is an element of the set of false propositions", or simply: ``$\varphi$ is false". Further properties of propositions can be expressed in a similar way. In \cite{lewsl} we have expressed knowledge of agents in this way, and the logic presented in \cite{lewjlc} has formulas of the form $\varphi:valid$ expressing that $\varphi$ belongs to the set of valid formulas.\footnote{In order to deal with a \textit{global predicate} such as validity, the non-Fregean logic developed in \cite{lewjlc} has necessarily the property that an equation $\varphi\equiv\psi$ is true in a model iff it is true in all models. This implies that we can identify a proposition with the same set of formulas in every model. Hence, it makes sense to speak of ``valid propositions".} 

Taking into account the axiom $(t:\varphi)\rightarrow\varphi$ from Justification Logic, together with the Tarski biconditionals $\varphi\leftrightarrow (\varphi:true)$, it follows that a justification is given by a set of true propositions $REASON_t\subseteq TRUE$. If $REASON_t$ is the set of true propositions assigned to the justification term $t$, then we put $NECESSARY:=\bigcup_{t\in Tm(C)} REASON_t$, where $Tm(C)$ is the set of all justification terms over the set of constant symbols $C$. $NECESSARY$ is the set of all propositions which are necessarily true in the model, and this set is the union of all justifications. That is, the concept of \textit{necessity} relies on the concept of \textit{justification}. In particular, the connection axiom $(t:\varphi)\rightarrow\square\varphi$ of logic $S4LP$ \cite{artnog} is satisfied. If a justification is a set of propositions, then it seems to be natural to read the symbol $:$ as ``is element of" in the same way as in formulas $\varphi:true$ or $\varphi:false$. Therefore, we write $\varphi:t$ instead of $t:\varphi$.\footnote{We hope not to annoy people working in the area of Justification Logic with this change of notation in this paper. We tried to use a mix of both notations: $t:\varphi$ and $\varphi:true$ which results in formulas such as $t:(\varphi:true)$. This seems to be not really satisfactory.} $\varphi:t$ reads as ``the proposition denoted by $\varphi$ is an element of the justification denoted by $t$". Shorter: ``$\varphi$ belongs to $t$" or simply ``$\varphi$ is $t$" if we consider a justification $t$ as a property of propositions. This seems to be in accordance with the ontological view presented in a recent article of Artemov \cite{art4} where justifications are regarded as sets of sentences.\footnote{Our interpretation of \textit{justifications as sets of true propositions} derives directly from the semantic approach presented in \cite{lewsl} (where \textit{knowledge} is given as a set of true propositions) and from axioms of basic Justification Logic independently from \cite{art4}. Connections between some properties of our non-Fregean semantics and the notion of \textit{modular model} given in \cite{art4} remain to be further investigated. We observe here that the inclusions in condition (2) [section 3, \cite{art4}] correspond in some sense to the equations of the homomorphism property of our Gamma-function: $\varGamma(s,\gamma)\cdot^\Lambda\varGamma(t,\gamma)=\varGamma(s\cdot t,\gamma)$ and $\varGamma(s,\gamma)+^\Lambda\varGamma(t,\gamma)=\varGamma(s+t,\gamma)$, where $\cdot^\Lambda$ and $+^\Lambda$ are operations of an algebra $\Lambda$ of (names of) justifications.} 
Justifications can be ordered by inclusion. We express this by formulas of the form $s\le t$. We deal with propositional quantifiers $\forall x$, $\exists x$ as in \cite{lewigpl}. In the present paper, we introduce justification quantifiers $\bigwedge u$ (``for all justifications $u$ ...") and $\bigvee u$ (``there is an justification $u$ ..."). To this end, our models have an additional universe consisting of elements called indexes which are in one-to-one correspondence with those subsets of the propositional universe which are justifications. The justification quantifiers then range over the universe of indexes which can be seen as (unique) names for justifications. This enables us to formulate our key axiom: $\square\varphi\leftrightarrow\bigvee u.(\varphi:u)$ which reads as ``$\varphi$ is necessarily true iff there exists a justification for $\varphi$". Necessity is a property of propositions which is first-order definable by means of justifications: consider the term $\{x\mid \bigvee u.(x:u)\}$, where $x$ is a propositional variable and $u$ is a variable for justifications.

Finally, we study extensions of our deductive system and of our semantics by some modal logic principles. We prove that our semantics is strong enough to restore the modal logics S4 and S5. These results can be proved independently from Justification Logic (instead of the axiom of application work with the axiom $K$ of modal logic). That is, we are able to establish a non-Fregean semantics of a logic that contains the modal system S4 (or S5). \\

We refer to the class of logics developed in this paper as $\in_J$-Logic (Epsilon-J-Logic). $J$ stands for ``justification", and the symbol $\in$ refers to the fact that we read a formula $\varphi:t$ as ``$\varphi$ is element of justification $t$".\footnote{This harmonizes with the names for $\in_T$-Logic \cite{zei,str,lewigpl} ($T$ stands for classical truth), $\in_I$-Logic \cite{lewnd,lewjlc} ($I$ stands for intuitionistic truth) and $\in_K$-Logic \cite{lewsl} ($K$ stands for knowledge).}

\section{Syntax}

$\in_J$-Logic extends the first-order non-Fregean logic studied in \cite{lewigpl}, which in turn is an extension of $\in_T$-Logic \cite{str, zei}. We have now two sorts of variables. $V_P=\{x_0,x_1,...\}$ is the well-ordered set of propositional variables, $V_J=\{v_0,v_1,...\}$ is the well-ordered set of justification variables. We refer to propositional variables as $x,y,z,x_1, ... $ and to justification variables as $u,v, u_1, ...$. $V:=V_P\cup V_J$ is the set of all variables. $C$ is a set of constant symbols for justifications and $D$ is a set of constant symbols for propositions. All these sets are pairwise disjoint. If there is no risk of confusion we may refer to the elements of $V\cup C\cup D$ as $x,y,x_1, ... $ without distinction of the actual sort of the element. The existential quantifiers $\exists x$ and $\bigvee u$ are definable by means of the universal quantifiers in the usual way. We use $\diamondsuit\varphi$ as an abbreviation for the formula $\neg\square\neg\varphi$. Besides propositional identity expressed by the identity connective $\equiv$ coming from basic non-Fregean logic we have an additional identity for justifications for which we use the same symbol: $s\equiv t$ reads as ``the justification terms $s$ and $t$ denote the same justification". We have a connective $<$ for propositional reference introduced and studied in \cite{lewman, lewnd, lewigpl}. $\varphi<\psi$ reads as ``the proposition denoted by $\psi$ says something about (refers to) the proposition denoted by $\varphi$". We introduce a relation symbol $\le$ for a partial ordering on justifications. The formula $s\le t$ reads as ``justification $s$ is contained in justification $t$" or ``justification $s$ is stronger than justification $t$". The predicates for truth and falsity come from $\in_T$-Logic \cite{str} and are already discussed in the introductory section.\footnote{Note that from a strict semantic point of view $:true$ is not a predicate but rather a predicate symbol which is interpreted in every model by the truth predicate $TRUE$ -- the set of true propositions of the model-theoretic universe. However, in the sense of Tarski's truth theory, where sentences (instead of propositions) are considered as the bearers of truth values, $:true$ can be seen as the truth predicate of the object language.}   

\begin{definition}\label{200}
Let $C, D$ be disjoint sets of constant symbols.
\begin{itemize}
\item The set of justification terms $Tm(C)$ is the smallest set that contains $V_J\cup C$ and is closed under the following condition. If $s,t\in Fm(C)$, then $(s\cdot t), (s+t)\in Tm(C)$. 
\item The set of formulas $Fm(C,D)$ is the smallest set that contains $V_P\cup D$ and is closed under the following condition. If $\varphi, \psi\in Fm(C,D)$, $s, t\in Tm(C)$, $x\in V_P$ and $u\in V_J$, then $(\varphi\rightarrow\psi)$, $(\neg\varphi)$, $(\varphi\equiv\psi)$, $(\varphi<\psi)$, $(s\equiv t)$, $(s\le t)$, $(\varphi:true)$, $(\varphi:false)$, $(\square\varphi)$, $(\varphi:t)$, $(\forall x.\varphi)$, $(\bigwedge u.\varphi$) are formulas.
\end{itemize}
\end{definition}

Let $\varphi\in Fm(C,D)$. Notions such as subformula of $\varphi$ and free variables of $\varphi$ are defined in the usual way. $\psi$ is a proper subformula of $\varphi$ if $\psi$ is a subformula of $\varphi$ and $\psi\neq\varphi$. We denote the set of all free propositional variables of $\varphi$ by $fvar_P(\varphi)$, the set of all free justification variables of $\varphi$ by $fvar_J(\varphi)$. We put $fvar(\varphi):=fvar_P(\varphi)\cup fvar_J(\varphi)$. $con_P(\varphi)$, $con_J(\varphi)$ is the set of all propositional constants, of all justification constants, occurring in $\varphi$, respectively. By $fcon(\varphi)$ we denote the set of all free variables and all constant symbols occurring in $\varphi$. If $t\in Tm(C)$, then $var(t)$, $con(t)$, is the set of variables, constant symbols occurring in $\varphi$, respectively. We put $varcon(t):=var(t)\cup con(t)$.

According to Definition \ref{200}, if $\varphi$ is a formula with $x,u\notin fvar(\varphi)$, then also $\forall x.\varphi$ and $\bigwedge u.\varphi$ are formulas, for example, $\forall x. d$, $\forall x. (y\rightarrow z)$, $\bigwedge u.(v\le v)$. Such formulas do not express meaningful propositions and are therefore undesired. Let us call a formula $\varphi$ proper if it has the following property: if $\psi$ is a subformula of $\varphi$ of the form $\forall x.\psi'$ or $\bigwedge u.\psi'$, then $x\in fvar_P(\psi')$, $u\in fvar_J(\psi')$, respectively. It is possible to modify and to extend Definition \ref{200} in such a way that $Fm(C,D)$ contains exactly the proper formulas (see \cite{lewigpl}). So we will assume in the following that $Fm(C,D)$ is the set of all proper formulas. That is, if we consider formulas $\forall x.\varphi$ or $\bigwedge u.\psi$, then we always assume that $x\in fvar_P(\varphi)$, $u\in fvar_J(\psi)$. Given the notion of free variables, it is clear that the set of proper formulas can be defined inductively and we therefore may carry out proofs by induction on the construction of (proper) formulas.

\begin{definition}\label{220}
A substitution is a function $\sigma:V_P\cup V_J\cup C\cup D\rightarrow Form(C,D)\cup Tm(C)$ with the property 
\begin{itemize}
\item $\sigma(x)\in Form(C,D)$ whenever $x\in V_P\cup D$
\item $\sigma(x)\in Tm(C)$ whenever $x\in V_J\cup C$
\end{itemize}
If $A\subseteq V_P\cup V_J\cup C\cup D$ and $\sigma$ is a substitution such that $\sigma(x)=x$ for all $x\in (V_P\cup V_J\cup C\cup D)\smallsetminus A$, then we write $\sigma:A\rightarrow Form(C,D)\cup Tm(C)$.
If $\sigma$ is a substitution, $x_0, ..., x_n\in V_P\cup V_J\cup C\cup D$ and $e_0, ..., e_n\in Form(C,D)\cup Tm(C)$ such that $x_i$ is a formula iff $e_i$ is a formula (equivalently: $x_i$ is a term iff $e_i$ is a term), then $\sigma [x_0:=e_0,...,x_n:=e_n]$ is the substitution $\tau$ defined as follows:
\begin{equation*}
\begin{split}
\tau(z)=
\begin{cases}
e_i &\text{ if }z=x_i, \text{ for some }i\le n\\
\sigma(z) &\text{ else.}
\end{cases}
\end{split}
\end{equation*}
The identity substitution $x\mapsto x$ is denoted by $\varepsilon$. Instead of $\varepsilon[x_0:=e_0,...,x_n:=e_n]$ we write $[x_0:=e_0,...,x_n:=e_n]$. A substitution $\sigma$ extends in the canonical way to a function $[\sigma]: Fm(C,D)\cup Tm(C)\rightarrow Fm(C,D)\cup Tm(C)$ (we use postfix notation for $[\sigma]$):
\begin{equation*}
\begin{split}
&x[\sigma]=\sigma(x), \text{ if } x\in V_P\cup V_J\cup C\cup D \\
&(s\cdot t)[\sigma]=(s[\sigma]\cdot t[\sigma])\\
&(s + t)[\sigma]=(s[\sigma] + t[\sigma])\\
&(\varphi\rightarrow\psi )[\sigma]=\varphi [\sigma]\rightarrow\psi [\sigma]\\
&(\neg\varphi)[\sigma]=\neg\varphi [\sigma]\\
&(\varphi:true)[\sigma]=\varphi [\sigma]:true\\
&(\varphi:false)[\sigma]=\varphi [\sigma]:false\\
&(\varphi:t)[\sigma]=\varphi[\sigma]:t [\sigma]\\
&(\square\varphi)[\sigma]=\square\varphi[\sigma]\\
&(s\equiv t )[\sigma]=s [\sigma]\equiv t [\sigma]\\
&(s\le t )[\sigma]=s [\sigma]\le t [\sigma]\\
&(\varphi\equiv\psi )[\sigma]=\varphi [\sigma]\equiv\psi [\sigma]\\
&(\varphi <\psi )[\sigma]=\varphi [\sigma] <\psi [\sigma]\\
&(\bigwedge u.\varphi)[\sigma]=\bigwedge v.\varphi[\sigma[u:=v]]\\
&(\forall x.\varphi)[\sigma]=\forall y.\varphi[\sigma[x:=y]],
\end{split}
\end{equation*}
where $v$ is the least variable of $V_J$ greater than all elements of $\bigcup\{fvar_J(\sigma(w))\mid w\in fcon(\bigwedge u.\varphi)\}$, and $y$ is the least variable of $V_P$ greater than all elements of $\bigcup\{fvar_P(\sigma(w))\mid w\in fcon(\bigwedge u.\varphi)\}$. We say that the variable $v$ (the variable $y$) is forced by the substitution $\sigma$ w.r.t. $\bigvee u.\varphi$ (w.r.t. $\forall x.\varphi$).
\end{definition}

The composition of two substitutions $\sigma$ and $\tau$ is the substitution $\sigma\circ\tau$ defined by $x\mapsto \sigma(z)[\tau]$. The following Lemma collects some useful properties of substitutions.

\begin{lemma}[Properties of substitutions]\label{240}
Let $\varphi\in Fm(C,D)$, $t\in Tm(C)$, and let $\sigma,\tau,\varrho$ be substitutions. Then
\begin{itemize}
\item $fcon(\varphi[\sigma])=\bigcup\{fcon(\sigma(y))\mid y\in fcon(\varphi)\}$ and \\
$varcon(t [\sigma])=\bigcup\{varcon(\sigma(y))\mid y\in varcon(t)\}$
\item If $\sigma(x)=\tau(x)$ for all $x\in fcon(\varphi)\cap varcon (t)$, then $\varphi[\sigma]=\varphi[\tau]$ and $t [\sigma]= t [\tau]$
\item The variable $y\in V_P$ forced by $\sigma$ w.r.t. $\forall x.\psi$ is the least element of $V_P$ greater than all elements of $fvar_P((\exists x.\psi)[\sigma])$
\item The variable $v\in V_J$ forced by $\sigma$ w.r.t. $\bigwedge u .\psi$ is the least element of $V_J$ greater than all elements of $fvar_J((\bigwedge u.\psi)[\sigma])$
\item $\varphi[\sigma\circ\tau]=\varphi[\sigma][\tau]$ and  $t [\sigma\circ\tau]= t [\sigma][\tau]$
\item $\sigma\circ (\tau\circ\delta)=(\sigma\circ\tau)\circ\delta$
\end{itemize}
\end{lemma}

\begin{definition}\label{260}
The alpha-congruence $=_\alpha$ is the smallest equivalence relation on $Form(C,D)$ satisfying the following conditions.
\begin{itemize}
\item $\varphi_1=_\alpha\psi_1$ and $\varphi_2=_\alpha\psi_2$ implies $\varphi_1 * \varphi_2=_\alpha \psi_1 * \psi_2$, where $*\in\{\rightarrow,\equiv,<\}$  
\item $\varphi=_\alpha\psi$ implies $\varphi:true=_\alpha\psi:true$, $\varphi:false=_\alpha\psi:false$, $\varphi:t=_\alpha\psi:t$, $\neg\varphi=_\alpha\neg\psi$, $\square\varphi=_\alpha\square\psi$
\item If $\bigwedge u.\varphi$ and $\bigwedge v.\psi$ are formulas\footnote{Recall that we mean \textit{proper formulas}, i.e. $u\in fvar_J(\varphi)$ and $v\in fvar(\psi)$.} such that $\varphi=_\alpha\psi[v:=u]$, and $v\neq u$ implies $u\notin fvar_J(\psi)$, then $\bigwedge u.\varphi=_\alpha\bigwedge v.\psi$.
\item If $\forall x.\varphi$ and $\forall y.\psi$ are formulas such that $\varphi=_\alpha\psi[y:=x]$, and $y\neq x$ implies $x\notin fvar(\psi)$, then $\forall x.\varphi=_\alpha\forall y.\psi$.
\end{itemize}         
\end{definition}      

Two formulas are alpha-congruent iff they differ at most on their bound variables. Applying the identity substitution $\varepsilon$ to a formula $\varphi$ results in general in a renaming of bound variables. $\varphi[\varepsilon]$ is in a certain \textit{normal form}, we say that it is normalized. It holds that $\varphi[\varepsilon]=_\alpha\varphi$, and furthermore $\varphi=_\alpha\psi$ iff $\varphi[\varepsilon]=\psi[\varepsilon]$ (see \cite{lewigpl} for proofs and more details). Our definition of semantics will ensure that two alpha-congruent formulas always denote the same proposition. From \cite{lewman, lewigpl} we adopt a further syntactical relation $\prec$ on the set of formulas. $\varphi\prec\psi$ will capture the intuitive notion of ``formula $\psi$ says something about formula $\varphi$" or ``formula $\psi$ refers to formula $\varphi$". 

\begin{definition}\label{280}
Let $\varphi,\psi\in Fm(C,D)$. Then $\varphi\prec\psi :\Leftrightarrow$ there are $x\in V_P$ and $\psi'\in Fm(C,D)\smallsetminus \{x\}$ such that $x\in fvar_P(\psi')$ and $\psi'[x:=\varphi]=_\alpha\psi$. The relation $\prec$ is called
syntactical reference.
\end{definition}

$\varphi\prec\psi$ implies that $\varphi$ is alpha-congruent to a proper subformula of $\psi$. In a quantifier-free language the converse would be true, too. However, $x\nprec\forall x. (x\rightarrow d)$, whereas $x\prec (x\rightarrow d)$. In the latter case, the formula $x\rightarrow d$ says something about formula $x$, namely that formula $x$ implies formula $d$. In the former case, however, the formula $\forall x. (x\rightarrow d)$ does not say anything about the formula $x$. 

A syntactical reference $\varphi\prec\psi$ can never be a self-reference since no formula is a proper subformula of itself: $\varphi\prec\varphi$ is impossible. There are no self-referential formulas. Self-reference must be shifted to the semantic level where it is represented by the semantic reference relation $<^\mathcal{M}$ on the propositional universe of a model $\mathcal{M}$. Our model definition ensures that syntactical reference $\varphi\prec\psi$ implies semantic reference between the respectively denoted propositions. This can be expressed by $\varphi<\psi$. If $\varphi\prec\psi$ and the formula $\varphi\equiv\psi$ is true in a given model, then the formula $\varphi <\varphi$ is also true in the model. Thus, $\varphi$ denotes a self-referential proposition. A typical example is the equation $d\equiv (d:true)$ asserting a truth teller. 
Constructions of models that contain specific self-referential propositions are developed in \cite{lewnd, lewsl}. For these constructions, however, we worked with a quantifier-free language. Constructions of infinite standard models for a first-order language are difficult because of the impredicativity of quantifiers (see \cite{lewigpl}). A standard model $\mathcal{M}$ is a model where every proposition is denoted by a sentence (i.e., there are no non-standard elements) and for any two formulas $\varphi,\psi$ and any assignment $\gamma$, $(\mathcal{M},\gamma)\vDash\varphi<\psi$ implies the existence of formulas $\varphi'$ and $\psi'$ such that $\varphi'\prec\psi'$ and $(\mathcal{M},\gamma)\vDash (\varphi'\equiv\varphi)\wedge (\psi'\equiv\psi)$, in particular $(\mathcal{M},\gamma)\vDash (\varphi' <\psi')$ (see \cite{lewigpl}). Standard models are the intended models. All models constructed in this article are standard models. The existence of non-standard models is the prize that we have to pay for the existence of a complete calculus. In \cite{lewigpl} we constructed a canonical model $\mathcal{M}$, i.e. a model without non-standard elements and with the following property: $\mathcal{M}\vDash\varphi<\psi\Leftrightarrow\varphi\prec\psi$. A canonical model is a standard model which satisfies only the trivial equations between sentences, i.e. equations between alpha-congruent sentences. One can modify the construction given in \cite{lewigpl} in such a way that the resulting standard model satisfies specific non-trivial equations. In this way, one gets standard models that contain specific self-referential propositions. 

We skip the proofs of the following useful facts.

\begin{lemma}\label{320}
If $\varphi\prec\psi$ and $\sigma$ is a substitution, then $\varphi[\sigma]\prec\psi[\sigma]$.
\end{lemma}

\begin{lemma}\label{340}
The syntactical reference $\prec$ is a transitive relation on $Fm(C,D)$.
\end{lemma}
            
Before we present our set of axioms we introduce the following notation. Let $\varphi\in Fm(C,D)$, $t\in Tm(C)$, let $\sigma,\sigma':V\rightarrow Fm(C,D)\cup Tm(C)$ be substitutions, and suppose $fvar(\varphi)=\{x_1,...,x_n\}$ and $var(t)=\{u_1,...,u_m\}$. Then we abbreviate the formula $(\sigma(x_1)\equiv\sigma'(x_1))\wedge ... \wedge (\sigma(x_n)\equiv\sigma'(x_n))$ by the notation $\sigma\equiv_\varphi\sigma'$ which can be informally read as ``$\sigma$ and $\sigma'$ coincide on all free variables of $\varphi$". Similarly, we write $\sigma\equiv_t\sigma'$ for the formula $(\sigma(u_1)\equiv\sigma'(u_1))\wedge ... \wedge (\sigma(u_m)\equiv\sigma'(u_m))$.\footnote{We consider the underlying orderings on $V_P$ and on $V_J$.} Note that $var(t)$ contains only justification variables whereas $fvar(\varphi)$ can contain justification variables as well as propositional variables.

\begin{definition}\label{380}
The set $Ax$ of axioms is the smallest set containing a sufficient set of tautologies of classical propositional logic, all formulas of the form given in (i) -- (xxiii) below, and being closed under the following two conditions:
\begin{itemize}
\item If $\varphi\in Ax$ and $u\in fvar_J(\varphi)$, then $\bigwedge u.\varphi\in Ax$. 
\item If $\varphi\in Ax$ and $x\in fvar_P(\varphi)$, then $\forall x.\varphi\in Ax$. 
\end{itemize}
\begin{enumerate}
\item $\varphi:true\leftrightarrow\varphi$ (\textit{Tarski biconditionals})
\item $\varphi:false\leftrightarrow\neg\varphi$
\item $(\varphi\rightarrow\psi ):s \rightarrow (\varphi:t\rightarrow\psi: (s\cdot t))$ (\textit{application})
\item $\varphi:s\rightarrow \varphi:(s+t)$ (\textit{weakening})
\item $\varphi:t\rightarrow \varphi:(s+t)$ (\textit{weakening})
\item $\square\varphi\leftrightarrow\bigvee u.(\varphi:u)$, if $u\in V_J\smallsetminus fvar_J(\varphi)$ (\textit{necessity as justified truth})
\item $\square\varphi\rightarrow\varphi$
\item $\varphi <\psi$, whenever $\varphi\prec\psi$ (\textit{syntactical reference implies semantical reference})
\item $(\varphi <\psi)\rightarrow ((\psi <\chi)\rightarrow (\varphi <\chi))$ (\textit{semantical reference is transitive})
\item $\varphi\equiv\psi$, whenever $\varphi=_\alpha\psi$ (\textit{alpha-congruent expressions have the same denotation})
\item $(\varphi\equiv\psi)\rightarrow (\varphi\rightarrow\psi)$
\item $(\sigma\equiv_\varphi\sigma')\rightarrow (\varphi[\sigma]\equiv\varphi[\sigma'])$ (\textit{Substitution Principle 1}) 
\item $\varphi[u:=t]\rightarrow \bigvee u.\varphi$ 
\item $\bigwedge u.\varphi\rightarrow\varphi[u:=t]$
\item $\bigwedge u.(\psi\rightarrow\varphi)\rightarrow (\bigwedge u.\psi\rightarrow\bigwedge u.\varphi)$
\item $\bigwedge u.(\psi\rightarrow\varphi)\rightarrow(\psi\rightarrow\bigwedge u.\varphi)$, if $u\notin fvar_J(\psi)$
\item $\varphi[x:=\psi]\rightarrow \exists x.\varphi$
\item $\forall x.\varphi\rightarrow\varphi[x:=\psi]$
\item $\forall x.(\psi\rightarrow\varphi)\rightarrow (\forall x.\psi\rightarrow\forall x.\varphi)$
\item $\forall x.(\psi\rightarrow\varphi)\rightarrow(\psi\rightarrow\forall x.\varphi)$, if $x\notin fvar_P(\psi)$
\item $(s\le t)\leftrightarrow \forall x. (x:s\rightarrow x:t)$
\item $(s\equiv t)\leftrightarrow (s\le t)\wedge (t\le s)$ (\textit{Extensionality: a justification is determined by its elements})
\item $(\sigma\equiv_t\sigma')\rightarrow (t[\sigma]\equiv t[\sigma'])$ (\textit{Substitution Principle 2}).
\end{enumerate}
\end{definition}

\begin{definition}\label{440}
If $\Phi\subseteq Fm(C,D)$, then $\Phi^\vdash$ is the smallest set of formulas containing $\Phi\cup Ax$ and being closed under the rule of Modus Ponens: If $\varphi\rightarrow\psi\in\Phi^\vdash$ and $\varphi\in\Phi^\vdash$, then $\psi\in\Phi^\vdash$.\\ 
If $\varphi\in \Phi^\vdash$, then we write $\Phi\vdash\varphi$. The notions of derivation, consistent set, inconsistent set, maximally consistent set are defined in the usual way.
\end{definition}

Notice that the axioms (xiii) and (xvii) derive from (xiv) and (xviii) together with propositional logic. The system $Ax$ is rather weak. No formula of the form $\varphi:t$ or $\square\varphi$ is a theorem. This will change when we add the rule of Axiom Necessitation in section 5. We will show that adding this rule together with the schema $\square\varphi\rightarrow\square\square\varphi$ will imply the modal logic principle of Necessitation (Theorem \ref{760}).

It is not hard to check that our deductive system extends Suszko's basic non-Fregean logic SCI, \textit{the Sentential Calculus with Identity}, (see, e.g., \cite{blosus}). For instance, the axiom $(\varphi\equiv\psi)\rightarrow(\neg\varphi\equiv\neg\psi)$ of SCI can be obtained as follows: Let $\sigma=[x:=\varphi]$, $\sigma'=[x:=\psi]$ be substitutions and let $\chi$ be the formula $\neg x$. Obviously, $(\sigma\equiv_\chi\sigma')\rightarrow (\chi[\sigma]\equiv\chi[\sigma'])$ is an instance of (xii), i.e. an element of $Ax$. But this notation is an abbreviation of the formula $(\varphi\equiv\psi)\rightarrow (\neg\varphi\equiv\neg\psi)$.

Our deductive system differs in some essential aspects from the axiomatization of (the weaker) first-order $\in_T$-Logic presented in \cite{zei}. For instance, we deal with substitutions in a different way and -- inspired by the Hilbert-style calculus for classical first-order logic given in \cite{rau} -- we work without a Generalization Rule.   

\section{Semantics}

Our notion of model extends the definition presented in \cite{lewigpl}. Instead of defining algebraic models in the style of Bloom/Suszko \cite{blosus} we prefer to work with an (equivalent) $\in_T$-style semantics which in its essence goes back to Str\"ater \cite{str} and Zeitz \cite{zei}. We add some algebraic structure in order to deal with justifications.   

\begin{definition}\label{460}
A model
\begin{equation*}
\mathcal{M}=(\varLambda,M,TRUE,NECESSARY,(REASON_l)_{l\in L},FALSE, <^\mathcal{M}, \varGamma)
\end{equation*}
is given by the following:
\begin{itemize}
\item $\varLambda=(L, +^\varLambda, \cdot^\varLambda,  \le^\varLambda)$ is an algebraic structure with two binary operations $+^\varLambda, \cdot^\varLambda$ and a partial ordering $\le^\varLambda$ on $L$. The elements of $L$ are called indexes or justification names.
\item $M$, $TRUE$, $FALSE$, $NECESSARY$, $REASON_l$, for each $l\in L$, are sets of propositions such that the following conditions are satisfied:
\begin{itemize}
\item $M=TRUE\cup FALSE$ is the propositional universe
\item $TRUE\cap FALSE=\varnothing$
\item $NECESSARY\subseteq TRUE$
\item $NECESSARY=\bigcup_{l\in L} REASON_l$
\item $(L,\le^\varLambda)$ is order-isomorphic to $((REASON_l)_{l\in L},\subseteq)$
\end{itemize}
\item $<^\mathcal{M}\subseteq M\times M$ is a transitive relation, called the reference relation of $\mathcal{M}$.
\item $\varGamma: (Fm(C,D)\cup Tm(C))\times (M\cup L)^{V}\rightarrow M$ is a semantic function, the so-called Gamma-function, that maps a formula $\varphi$ to a proposition $\varGamma(\varphi,\gamma)\in M$, and a justification term $s$ to an index $\varGamma(s,\gamma)\in L$. $\varGamma$ depends on assignments $\gamma:V\rightarrow M\cup L$. If $\gamma$ is an assignment, $x\in V_P$ and $m\in M$, then $\gamma_x^m$ is the assignment that maps $x$ to $m$ and variables $y\in V\smallsetminus\{x\}$ to $\gamma(y)$. The assignment $\gamma_u^l$, where $u\in V_J$ and $l\in L$, is defined similarly.  
\end{itemize}
The Gamma-function satisfies the following \textbf{structure conditions}:
\begin{itemize}
\item For all $x\in V=V_P\cup V_J$ and all assignments $\gamma$: $\varGamma(x,\gamma)=\gamma(x)$. (Extension Property (EP))
\item If $\gamma(x)=\gamma'(x)$ for all $x\in fvar(\varphi)$ and $\gamma(u)=\gamma'(u)$ for all $u\in var_J(s)$, then $\varGamma(\varphi, \gamma)=\varGamma(\varphi,\gamma')$ and $\varGamma(s,\gamma)=\varGamma(s,\gamma')$. (Coincidence Property (CP))\footnote{If $fvar(\varphi)=\varnothing$, then (CP) justifies to write $\varGamma(\varphi)$ instead of $\varGamma(\varphi,\gamma)$.} 
\item If $\sigma:V\rightarrow Fm(C,D)\cup Tm(C)$ is a substitution and $\gamma:V\rightarrow M\cup L$ an assignment, then $\varGamma(\varphi[\sigma],\gamma)=\varGamma(\varphi,\gamma\sigma)$ and $\varGamma(s[\sigma],\gamma)=\varGamma(s,\gamma\sigma)$, where $\gamma\sigma:V\rightarrow Fm(C,D)\cup Tm(C)$ is the substitution defined by $x\mapsto\varGamma(\sigma(x),\gamma)$ for $x\in V=V_P\cup V_J$. (Substitution Property (SP))
\item $\varphi\prec\psi$ implies $\varGamma(\varphi,\gamma)<^\mathcal{M}\varGamma(\psi,\gamma)$, for all assignments $\gamma:V\rightarrow M\cup L$. (Reference Property (RP))
\item $\varGamma(s + t,\gamma)=\varGamma(s,\gamma) +^\varLambda\varGamma(t,\gamma)$ and $\varGamma(s\cdot t,\gamma)=\varGamma(s,\gamma)\cdot^\varLambda\varGamma(t,\gamma)$, for all assignments $\gamma\in (M\cup L)^V$. (Homomorphism Property (HP))
\end{itemize}
The Gamma-function satisfies the following \textbf{truth conditions}. For all assignments $\gamma:V\rightarrow M\cup L$, for all formulas $\varphi,\psi$, and for all justification terms $s,t$:
\begin{enumerate}
\item $\varGamma(\varphi\rightarrow \psi,\gamma)\in TRUE\Leftrightarrow\varGamma(\varphi,\gamma)\in FALSE\text{ or }\varGamma(\psi,\gamma)\in TRUE$
\item $\varGamma(\neg\varphi,\gamma)\in TRUE\Leftrightarrow\varGamma(\varphi,\gamma)\notin TRUE$
\item $\varGamma(\varphi:true,\gamma)\in TRUE\Leftrightarrow\varGamma(\varphi,\gamma)\in TRUE$
\item $\varGamma(\varphi:false,\gamma)\in TRUE\Leftrightarrow\varGamma(\varphi,\gamma)\in FALSE$
\item $\varGamma(\varphi\equiv\psi,\gamma)\in TRUE\Leftrightarrow\varGamma(\varphi,\gamma)=\varGamma(\psi,\gamma)$
\item $\varGamma(\varphi <\psi,\gamma)\in TRUE\Leftrightarrow\varGamma(\varphi,\gamma)<^\mathcal{M}\varGamma(\psi,\gamma)$
\item $\varGamma(s\equiv t,\gamma)\in TRUE\Leftrightarrow \varGamma(s,\gamma)=\varGamma(t,\gamma)$
\item $\varGamma(s\le t,\gamma)\in TRUE\Leftrightarrow\varGamma(s,\gamma)\le^\varLambda\varGamma(t,\gamma)$
\item $\varGamma(\square\varphi,\gamma)\in TRUE\Leftrightarrow\varGamma(\varphi,\gamma)\in NECESSARY$
\item $\varGamma(\varphi:t,\gamma)\in TRUE\Leftrightarrow\varGamma(\varphi,\gamma)\in REASON_{\varGamma(t,\gamma)}$
\item $\varGamma(\varphi\rightarrow \psi,\gamma)\in REASON_{\varGamma(s,\gamma)}$ and $\varGamma(\varphi,\gamma)\in REASON_{\varGamma(t,\gamma)}$ implies $\varGamma(\psi,\gamma)\in REASON_{\varGamma(s\cdot t,\gamma)}$
\item $\varGamma(\varphi,\gamma)\in REASON_{\varGamma(s,\gamma)}\cup REASON_{\varGamma(t,\gamma)}$ implies\\ $\varGamma(\varphi,\gamma)\in REASON_{\varGamma(s+t,\gamma)}$
\item $\varGamma(\bigwedge u.\varphi,\gamma)\in TRUE\Leftrightarrow\varGamma(\varphi,\gamma_u^l)\in TRUE$ for all $l\in L$
\item $\varGamma(\forall x.\varphi,\gamma)\in TRUE\Leftrightarrow\varGamma(\varphi,\gamma_x^m)\in TRUE$ for all $m\in M$
\end{enumerate}
If $\mathcal{M}$ is a model and $\gamma:V\rightarrow M\cup L$ is an assignment, then the tuple $(\mathcal{M},\gamma)$ is called an interpretation.
\end{definition}

\begin{definition}\label{470}
Let $(\mathcal{M},\gamma)$ be an interpretation. The satisfaction relation is defined by $(\mathcal{M},\gamma)\vDash\varphi :\Leftrightarrow\varGamma(\varphi,\gamma)\in TRUE$, for $\varphi\in Fm(C,D)$. If $\Phi\subseteq Fm(C,D)$, then $(\mathcal{M},\gamma)\vDash\Phi :\Leftrightarrow(\mathcal{M},\gamma)\vDash\varphi$ for every $\varphi\in \Phi$. An interpretation that satisfies a formula (a set of formulas) is called a model of that formula (of that set of formulas). The class of all interpretations that satisfy a set $\Phi$ is $Mod(\Phi):=\{(\mathcal{M},\gamma)\vDash\Phi\mid\mathcal{M}$ is a model and $\gamma$ is an assignment of $\mathcal{M}\}$. The relation of logical consequence is defined as follows: $\Phi\Vdash\varphi:\Leftrightarrow Mod(\Phi)\subseteq Mod(\{\varphi\})$.
\end{definition}

Let $\mathcal{M}$ be a model and $\gamma$ be an assignment. The set 
\begin{equation*}
F_N:=\{\varphi\in Fm(C,D)\mid\varGamma(\varphi,\gamma)\in NECESSARY\}
\end{equation*}
is closed under Modus Ponens (because of truth condition (xi), see the proof of Theorem \ref{820}) but in general it does not contain the axioms of $Ax$. So it is in general not closed under logical consequence. This will change when we introduce the rule (and the corresponding truth condition) of Axiom Necessitation ensuring that $F_N$ contains all axioms of $Ax$. \\

The structure conditions of a model guarantee the following Substitution Principles.

\begin{lemma}[Substitution Principles 1 and 2]\label{475}
Let $\sigma,\sigma':V\rightarrow Fm(C,D)\cup Tm(C)$ be substitutions and let $\varphi\in Fm(C,D)$ and $t\in Tm(C)$. Then
\begin{itemize}
\item $\Vdash(\sigma\equiv_\varphi\sigma')\rightarrow (\varphi[\sigma]\equiv\varphi[\sigma'])$ 
\item $\Vdash(\sigma\equiv_t\sigma')\rightarrow (t[\sigma]\equiv t[\sigma'])$
\end{itemize}
\end{lemma}

\paragraph*{Proof.}
Suppose $(\mathcal{M},\gamma)$ is a model of $\sigma\equiv_\varphi\sigma'$. This means that $\varGamma(\sigma(x),\gamma)=\varGamma(\sigma'(x),\gamma)$, for all $x\in fvar(\varphi)$. By (SP) and (EP), $\gamma\sigma(x)=\varGamma(x,\gamma\sigma)=\varGamma(x,\gamma\sigma')=\gamma\sigma'(x)$, for all $x\in fvar(\varphi)$. Then (CP) and (SP) yield $\varGamma(\varphi[\sigma],\gamma)=\varGamma(\varphi,\gamma\sigma)=\varGamma(\varphi,\gamma\sigma')=\varGamma(\varphi[\sigma'],\gamma)$. That is, $(\mathcal{M},\gamma)\vDash\varphi[\sigma]\equiv\varphi[\sigma']$. The second Substitution Principle follows similarly. $\square$\\

It is now straightforward to show that all axioms are valid.\\

Stronger than Substitution Principle 1 is item (i) of the following \textit{Substitution Lemma}. Several versions of this Lemma were already proved in \cite{zei,str}. The proof given in [Lemma 3.14, \cite{lewnd}] relies on ideas due to Zeitz \cite{zei} and can easily be adapted to $\in_J$-Logic.

\begin{lemma}[Substitution Lemma]\label{480}
Let $\mathcal{M}$ be a model, $\varphi\in Fm(C,D)$.
\begin{enumerate}
\item If $\sigma,\sigma'$ are substitutions and $\gamma,\gamma':V\rightarrow M\cup L$ are assignments such that $\varGamma(\sigma(z),\gamma)=\varGamma(\sigma'(z),\gamma')$ for all $z\in fcon(\varphi)$, then $\varGamma(\varphi[\sigma],\gamma)=\varGamma(\varphi[\sigma'],\gamma')$.
\item If $\gamma:V\rightarrow M\cup L$ is an assignment and $\sigma$ is a substitution such that $\varGamma(e)=\varGamma(\sigma(e),\gamma)$ for every $e\in C\cup D$, then $\varGamma(\varphi[\sigma],\gamma)=\varGamma(\varphi,\gamma\sigma)$.
\end{enumerate}
\end{lemma}

Item (ii) provides a condition such that the (SP) also holds for certain substitutions which are not necessarily restricted to the domain of variables. Item (i) implies a third Substitution Principle which we will apply to show that the following sets are well defined. Let $\mathcal{M}$ be a model. We define:
\begin{equation*}
\begin{split}
&POSSIBLE:=\{\varGamma(\varphi,\gamma)\mid\varGamma(\neg\varphi,\gamma)\notin NECESSARY,\text{ for some }\varphi,\gamma\}\\
&IMPOSSIBLE:=\{\varGamma(\varphi,\gamma)\mid\varGamma(\neg\varphi,\gamma)\in NECESSARY,\text{ for some }\varphi,\gamma\}
\end{split}
\end{equation*}

Suppose $\beta$ is an assignment and $\psi$ is a formula such that $\varGamma(\psi,\beta)=p\in POSSIBLE$. By definition of $POSSIBLE$, there is a formula $\varphi$ and an assignment $\gamma$ such that $\varGamma(\varphi,\gamma)=p$ and $\varGamma(\neg\varphi,\gamma)\notin NECESSARY$. We define two substitutions $\sigma=[x:=\varphi]$ and $\sigma'=[x:=\psi]$. Then $\varGamma(\sigma(x),\gamma)=\varGamma(\varphi,\gamma)=p=\varGamma(\psi,\beta)=\varGamma(\sigma'(x),\beta)$. Item (i) of the Substitution Lemma yields $\varGamma(\neg\varphi,\gamma)=\varGamma((\neg x)[\sigma],\gamma)=\varGamma((\neg x)[\sigma']),\beta)=\varGamma(\neg\psi,\beta)$. Thus, $\varGamma(\neg\psi,\gamma)\notin NECESSARY$. Hence, $POSSIBLE$ is well defined. Similarly, one shows that $IMPOSSIBLE$ is well defined.\\

The next fact follows readily from the definitions.

\begin{lemma}\label{482}
Let $\mathcal{M}$ be a model. For any $\varphi\in Fm(C,D)$ and any assignment $\gamma:V\rightarrow Fm(C,D)\cup Tm(C)$: 
\begin{itemize}
\item $\varGamma(\varphi,\gamma)\in POSSIBLE\Leftrightarrow\varGamma(\neg\square\neg\varphi,\gamma)\in TRUE\Leftrightarrow\varGamma(\diamondsuit\varphi,\gamma)\in TRUE$
\item $\varGamma(\varphi,\gamma)\in IMPOSSIBLE\Leftrightarrow\varGamma(\square\neg\varphi,\gamma)\in TRUE\Leftrightarrow\varGamma(\neg\diamondsuit\varphi,\gamma)\in TRUE$
\end{itemize}
\end{lemma}

Note that the set $FALSE$ is definable, too: $FALSE=\{\varGamma(\varphi,\gamma)\mid\varGamma(\neg\varphi,\gamma)\in TRUE,\text{ for some }\varphi, \gamma\}$. In fact, formulas of the form $\varphi:false$ and $\neg\varphi$ are logically equivalent and we could do without the operator $:false$. However, in this intensional setting $\neg\varphi$ and $\varphi:false$ express different intensions and denote in general different propositions.\\

As already pointed out above, we expect that any two alpha-congruent formulas denote the same proposition. This is actually the case as the following Lemma shows. We adopt the proof from \cite{lewigpl}.

\begin{lemma}[Alpha Property $(\alpha P)$]
Let $\mathcal{M}$ be a model. For all formulas $\varphi,\psi$ and all assignments $\gamma:V\rightarrow M$, if $\varphi=_\alpha\psi$, then $\varGamma(\varphi,\gamma)=\varGamma(\psi,\gamma)$.
\end{lemma}

\paragraph*{Proof.}
Suppose $\varphi=_\alpha\psi$. Recall that this is equivalent with the condition $\varphi[\varepsilon]=\psi[\varepsilon]$, where $\varepsilon$ is the identity substitution (see \cite{lewigpl}). It holds that $\gamma=\gamma\varepsilon$, for any assignment $\gamma:V\rightarrow M\cup L$. By (SP) of $\mathcal{M}$: $\varGamma(\varphi,\gamma)=\varGamma(\varphi,\gamma\varepsilon)=\varGamma(\varphi[\varepsilon],\gamma)=\varGamma(\psi[\varepsilon],\gamma)=\varGamma(\psi,\gamma\varepsilon)=\varGamma(\psi,\gamma)$. $\square$

\section{The Completeness Theorem}

In order to prove that our deductive system is complete with respect to the defined semantics we follow the usual strategy. We define a suitable notion of Henkin set, show that every Henkin set has a model, and finally show that every consistent set extends to a Henkin set. Of course, this establishes in particular soundness of a certain subset of $Ax$ with respect to the semantics of the particular (and less expressive) $\in_T$-logic presented in \cite{lewigpl}.

\begin{lemma}\label{485}
Let $\varphi\in Ax$. If $c\in C$ and $w\in V_J\smallsetminus var(\varphi)$, then $\varphi[c:=w]\in Ax$. Similarly, if $d\in C$ and $y\in V_P\smallsetminus var(\varphi)$, then $\varphi[d:=y]\in Ax$. 
\end{lemma}

\paragraph*{Proof.}
Consider, for instance, the axiom $\bigwedge u.\varphi\rightarrow\varphi[u:=t]$ of schema (xiv). We have 
\begin{equation*}
\begin{split}
&(\bigwedge u.\varphi\rightarrow\varphi[u:=t])[c:=w]\\
&=(\bigwedge u.\varphi)[c:=w]\rightarrow(\varphi[u:=t])[c:=w]\\
&= \bigwedge v.\varphi[c:=w,u:=v]\rightarrow\varphi[c:=w] [u:=t']\\
&=\bigwedge v.\varphi[c:=w] [u:=v]\rightarrow (\varphi[c:=w] [u:=v])[v:=t']\\
&=\bigwedge v.\psi\rightarrow \psi[v:=t'],
\end{split}
\end{equation*}
where $v$ is the variable forced by $[c:=w]$ w.r.t. $\bigwedge u.\varphi\rightarrow\varphi[u:=t]$, $t'=t[c:=w]$, and $\psi:=\varphi[c:=w] [u:=v]$. The last formula is clearly an axiom of the form (xiv). Note that $w\neq u$ since $w\notin var(\varphi)$. The same argument can also be applied in the other cases that involve quantifiers. The quantifier-free cases follow straightforwardly. $\square$

\begin{lemma}\label{490}
If $\Phi\vdash\varphi$ and $c\in con_J(\varphi)\smallsetminus con_J(\Phi)$, then for any $u\in V_J\smallsetminus var(\varphi)$, $\Phi\vdash\bigwedge u.(\varphi[c:=u])$. Similarly, if $\Phi\vdash\varphi$ and $d\in con_P(\varphi)\smallsetminus con_P(\Phi)$, then for any $x\in V_P\smallsetminus var(\varphi)$, $\Phi\vdash\forall x.(\varphi[d:=x])$. 
\end{lemma}

\paragraph*{Proof.}
The proof is an induction on the length $n\ge 1$ of a derivation $\Phi\vdash\varphi$. If $n=1$, then $\varphi$ must be an axiom (note that $\varphi\in\Phi$ is impossible because $c\in con(\varphi)$ does not occur in $\Phi$). By Lemma \ref{485}, $\varphi[c:=u]$ is an axiom, too. Thus, $\bigwedge u.(\varphi[c:=u])$ is an axiom. Now suppose $n>1$. We may assume that $\varphi$ is obtained by Modus Ponens, that is, $\Phi\vdash\psi\rightarrow\varphi$ and $\Phi\vdash\psi$, for some formula $\psi$. If $c\notin con_J(\psi)$, then $\psi\rightarrow\varphi[c:=u]=(\psi\rightarrow\varphi)[c:=u]$, where $u\in V_J\smallsetminus var(\psi)$. By the induction hypothesis, $\Phi\vdash\bigwedge u.(\psi\rightarrow\varphi[c:=u])$. By axiom (xvi), $\Phi\vdash\psi\rightarrow\bigwedge u.(\varphi[c:=u])$. Now we apply Modus Ponens. 
If $c\in con_J(\psi)$, then we may apply the induction hypothesis to $\Phi\vdash\psi$ and to $\Phi\vdash\psi\rightarrow\varphi$, and obtain $\Phi\vdash\bigwedge u.(\psi[c:=u])$ and $\Phi\vdash\bigwedge u.(\psi[c:=u]\rightarrow\varphi[c:=u])$. Axiom (xv) and Modus Ponens now yield the assertion. The second assertion follows analogously. $\square$\\

In our treatment of Henkin set (Definitions \ref{520} and \ref{540}, and Lemma \ref{560} below) we follow some notation and techniques given in W. Rautenberg's logic book \cite{rau}. 

\begin{definition}\label{520}
A set $\Phi\subseteq Fm(C,D)$ is called a Henkin set if 
\begin{itemize}
\item $\Phi$ is maximally consistent
\item $\Phi\vdash\bigwedge u.\varphi \Leftrightarrow \Phi\vdash\varphi[u:=c]$ for all $c\in C$
\item $\Phi\vdash\forall x.\varphi \Leftrightarrow \Phi\vdash\varphi[x:=d]$ for all $d\in D$
\end{itemize}
\end{definition}

\begin{definition}\label{540}
Suppose we are given the language $Fm(C,D)$.
\begin{itemize}
\item To each pair $\varphi,u$, where $\varphi\in Fm(C,D)$ and $u\in fvar_J(\varphi)$, we assign exactly one new justification constant $c_{\varphi,u}\notin C$ and define $\varphi^u:=\neg\bigwedge u.\varphi\wedge\varphi[u:=c_{\varphi,u}]$. Furthermore, $X(C,D):=\{\neg(\varphi^u)\mid \varphi\in Fm(C,D), u\in fvar_J(\varphi)\}$.
\item To each pair $\varphi,x$, where $\varphi\in Fm(C,D)$ and $x\in fvar_P(\varphi)$, we assign exactly one new propositional constant $d_{\varphi,x}\notin D$ and define $\varphi^x:=\neg\forall x.\varphi\wedge\varphi[x:=d_{\varphi,x}]$. Furthermore, $Y(C,D):=\{\neg(\varphi^x)\mid \varphi\in Fm(C,D), x\in fvar_P(\varphi)\}$.
\end{itemize}
\end{definition} 

Note that $\neg(\varphi^u)$ can be written as $\bigvee u.\neg\varphi\rightarrow\neg\varphi[u:=c_{\varphi,u}]$. In this sense, the new constant symbol $c_{\varphi,u}$ is a \textit{witness} for the truth of $\bigvee u.\neg\varphi$. Similarly, $c_{\varphi,x}$ can be seen as a witness for the truth of $\exists x.\neg\varphi$.

\begin{lemma}\label{560}
If $\Phi\subseteq Fm(C,D)$ is consistent, then $\Phi\cup X(C,D)\subseteq Fm(C',D)$ and $\Phi\cup Y(C,D)\subseteq Fm(C,D')$ are consistent, too, where $C'$, $D'$ are the sets $C$, $D$ enriched by the new constant symbols, respectively.
\end{lemma}

\paragraph*{Proof.}
Suppose $\Phi\cup X(C,D)\subseteq Fm(C',D)$ is inconsistent. There are formulas $\neg(\varphi_0^{u_0}), ...,\neg(\varphi_n^{u_n})\in X(C,D)$ such that $\Phi\cup\{\neg(\varphi_i^{u_i})\mid i\le n\}$ is inconsistent. We may assume that $n$ is minimal with this property. Let $u:=u_n$, $\varphi:=\varphi_n$, $c:=c_{n,\varphi}$, $\Phi':=\Phi\cup\{\neg(\varphi_i^{u_i})\mid i< n\}$. Then $\Phi'$ is consistent and $\Phi'\cup\{\neg(\varphi^u)\}$ is inconsistent. From classical propositional logic it follows that $\Phi'\vdash\varphi^u$. That is, $\Phi'\vdash\neg\bigwedge u.\varphi\wedge\varphi[u:=c]$, thus $\Phi'\vdash\neg\bigwedge u.\varphi$ and $\Phi'\vdash\varphi[u:=c]$. By construction, $c\in con_J(\varphi)\smallsetminus con_J(\Phi')$ and, of course, $u\notin var_J(\varphi[u:=c])$. Then we can apply Lemma \ref{490} and get $\Phi'\vdash\bigwedge u.\varphi$ and $\Phi'\vdash\neg\bigwedge u.\varphi$. It follows that $\Phi'$ is inconsistent. This contradiction shows that $\Phi\cup X(C,D)\subseteq Fm(C',D)$ is consistent. In a similar way, one shows the consistency of $\Phi\cup Y(C,D)\subseteq Fm(C,D')$. $\square$

\begin{definition}\label{580}
Let $\Phi\subseteq Fm(C,D)$ be a maximally consistent set. For $\varphi,\psi\in Fm(C,D)$ and $s,t\in Tm(C)$ define 
\begin{itemize}
\item $\varphi\approx_F\psi :\Leftrightarrow\Phi\vdash\varphi\equiv\psi$,
\item $s\approx_T t :\Leftrightarrow\Phi\vdash s\equiv t$.
\end{itemize} 
\end{definition}

Of course, these equivalence relations rely on the given set $\Phi$. It will be clear from the context on which maximally consistent set these relations are based.

\begin{lemma}\label{600}
If $\Phi\subseteq Fm(C,D)$ is maximally consistent, then:
\begin{enumerate}
\item $\approx_T$ is an equivalence relation on $Tm(C)$.
\item $\approx_F$ is an equivalence relation on $Fm(C,D)$, containing alpha-congruence.
\item If $\varphi\approx_F\varphi'$ and $\psi\approx_F\psi'$ and $s\in Tm(C)$, then 
\begin{equation*}
\begin{split}
\square\varphi &\approx_F\square\varphi'\\
\varphi:s &\approx_F\varphi':s\\
\varphi<\psi &\approx_F\varphi'<\psi'.
\end{split}
\end{equation*}
\item If $s\approx_T s'$ and $t\approx_T t'$ and $\varphi\in Fm(C,D)$, then 
\begin{equation*}
\begin{split}
\varphi:s &\approx_F\varphi:s'\\ 
s\le t &\approx_F s'\le t'\\
s+t &\approx_T s'+t'\\ 
s\cdot t &\approx_T s'\cdot t'.
\end{split}
\end{equation*}
\item If $\varphi\approx_F\psi$, then $\varphi\in\Phi\Leftrightarrow\psi\in\Phi$. 
\end{enumerate}
\end{lemma}

\paragraph*{Proof.}
(i): By Lemma \ref{490}, $\vdash (d:s)\rightarrow (d:s)$ implies $\vdash\forall x. (x:s\rightarrow x:s)$, where $s$ is a justification term and $d$ is any propositional constant. It follows that $\Phi\vdash\forall x.(x:s\rightarrow x:s)$, thus $s\approx_T s$ and $\approx_T$ is reflexive. Suppose $s_1\approx_T s_2$ and $s_2\approx_T s_3$. Then $\Phi\vdash\forall x.(x:s_1\rightarrow x:s_2)$ and $\Phi\vdash\forall x.(x:s_2\rightarrow x:s_3)$. Working with axiom (xix) and axioms of classical propositional logic we get $\Phi\vdash\forall x.(x:s_1)\rightarrow\forall x.(x:s_3)$ and thus $\Phi\vdash\forall x.(x:s_1\rightarrow x:s_3)$. Similarly, we obtain $\Phi\vdash\forall x.(x:s_3\rightarrow x:s_1)$. That is, $s_1\approx_T s_3$ and $\approx_T$ is transitive. One easily shows that $\approx_T$ is also symmetric.\\ 
(ii): From axiom (x) it follows that $\approx_F$ is reflexive and contains alpha-congruence. Suppose $\varphi\approx_F\psi$ and let $\chi:= (x\approx_F\varphi)$, where $x\notin var_P(\varphi)$. Then $\chi[x:=\varphi]\approx_F \chi[x:=\psi]$, by axiom (xii). Since $\Phi\vdash\chi[x:=\varphi]$, axiom (xi) yields $\Phi\vdash\chi[:=\psi]$. Thus, $\psi\approx_F\varphi$ and $\approx_F$ is symmetric. Now suppose $\varphi_1\approx_F\varphi_2$ and $\varphi_2\approx_F\varphi_3$. Let $\chi:= (x\equiv\varphi_3)$, with $x\notin var_P(\varphi_3)$. By axiom (xii), $\chi[x:=\varphi_1]\approx_F\chi[x:=\varphi_2]$. By hypothesis, $\Phi\vdash\chi[x:=\varphi_2]$. Symmetry of $\approx_F$, axiom (xi) and Modus Pones yield $\Phi\vdash\chi[x:=\varphi_1]$. That is, $\varphi_1\approx_F\varphi_3$ and $\approx_F$ is transitive.\\
(iii): By axiom (x), $\square x\approx_F\square x$, for any $x\in V_P$. By axiom (xii), $\square\varphi = (\square x)[x:=\varphi\approx (\square x)[x:=\varphi']=\square\varphi'$. The second assertion of (iii) follows similarly. Now suppose $\varphi\approx_F\varphi'$ and $\psi\approx_F\psi'$. Let $x\in V_P\smallsetminus var_P(\psi')$ and $y\in V_P\smallsetminus var_P(\varphi)$. By axiom (xii), $\varphi <\psi = (\varphi < y)[y:=\psi]\approx_F (\varphi < y)[y:=\psi'] = \varphi<\psi' = (x <\psi')[x:=\varphi]\approx_F (x <\psi')[x:=\varphi'] = \varphi' <\psi'$. Transitivity of $\approx_F$ yields $\varphi <\psi\approx_F\varphi'<\psi'$.\\
(iv) follows in a similar way as (iii).\\
(v) follows from axiom (xi) and symmetry of $\approx_F$.
$\square$

\begin{theorem}\label{620}
Every Henkin set has a model.
\end{theorem}

\paragraph*{Proof.}
Let $\Phi\subseteq Fm(C,D)$ be a Henkin set. We consider the equivalence relations $\approx_F$ and $\approx_T$ w.r.t. to the maximally consistent set $\Phi$. Let $\overline{\varphi}$ be the equivalence class of $\varphi\in Fm(C,D)$ modulo $\approx_F$, and let $\overline{s}$ be the equivalence class of $s\in Tm(C)$ modulo $\approx_s$.\\
\textbf{Claim} 1: For every $t\in Tm(C)$ there is a $c\in C$ such that $c\approx_T t$. For every $\varphi\in Fm(C,D)$ there is a $d\in D$ such that $d\approx_F \varphi$.\\
\textit{Proof of the Claim}: If $u\in V_J\smallsetminus var(t)$, then obviously $\Phi\vdash (u\equiv t)[u:=t]$. By axiom (xiii), $\Phi\vdash\bigvee u.(u\equiv t)$. That is, $\Phi\vdash\neg\bigwedge u.\neg (u\equiv t)$. Since $\Phi$ is consistent, $\Phi\nvdash\bigwedge u.\neg (u\equiv t)$. Since $\Phi$ is a Henkin set, $\Phi\nvdash\neg (c\equiv t)$ for some $c\in C$. By maximally consistency of $\Phi$, $\Phi\vdash c\equiv t$. Similarly for the second assertion. This proves Claim 1.\\
The ingredients of our model are now given as follows:
\begin{equation*}
\begin{split}
&M:=\{\overline{\varphi}\mid \varphi\in Fm(C,D)\}=\{\overline{d}\mid d\in D\}\\
&TRUE:=\{\overline{\varphi}\mid \varphi\in\Phi\}=\{\overline{\varphi}\mid \varphi:true\in\Phi\}\\ 
&FALSE:=\{\overline{\varphi}\mid \varphi\notin\Phi\}=\{\overline{\varphi}\mid \varphi:false\in\Phi\}=\{\overline{\varphi}\mid\neg\varphi\in\Phi\}\\
&NECESSARY:=\{\overline{\varphi}\mid\square\varphi\in\Phi\}\\
&REASON_{\overline{s}}:=\{\overline{\varphi}\mid \varphi:s\in\Phi\text{ and }s\in Tm(C)\}\\
&<^\mathcal{M}:=\{(\overline{\varphi},\overline{\psi})\mid \varphi<\psi\in\Phi\}\\
&L:=\{\overline{s}\mid s\in Tm(C)\}=\{\overline{c}\mid c\in C\}\\
&\overline{s}+^\varLambda\overline{t}:=\overline{s+t}\\
&\overline{s}\cdot^\varLambda\overline{t}:=\overline{s\cdot t}\\
&\le^\varLambda:=\{(\overline{s},\overline{t})\mid s\le t\in\Phi\}\\
&\varLambda:=(L,+^\varLambda,\cdot^\varLambda,\le^\varLambda)
\end{split}
\end{equation*}
It will be helpful to have in mind that equivalence classes $\overline{\varphi}$, $\overline{s}$ can be represented by $\overline{d}$, $\overline{c}$, for some $d\in D$, $c\in C$, respectively. 
By Lemma \ref{600}, all the above sets and operations are well-defined.\\
\textbf{Claim} 2: $NECESSARY=\bigcup_{\overline{s}\in L} REASON_{\overline{s}}$, and the map $\overline{s}\mapsto REASON_{\overline{s}}$ is an order-isomorphism from $(L,\le^\varLambda)$ to $((REASON_{\overline{s}})_{s\in Tm(C)},\subseteq)$.\\
\textit{Proof of the Claim}: If $\overline{\varphi}\in \bigcup_{\overline{s}\in L} REASON_{\overline{s}}$, then $\Phi\vdash\varphi:c$ for some $c\in C$. Choose $u\in V_J\smallsetminus fvar(\varphi)$. Then $\Phi\vdash(\varphi:u)[u:=c]$. By axiom (xiii), $\Phi\vdash\bigvee u.(\varphi:u)$. Axiom (vi) yields $\Phi\vdash\square\varphi$, thus $\overline{\varphi}\in NECESSARY$. Now let $\overline{\varphi}\in NECESSARY$, that is, $\Phi\vdash\square\varphi$. By axiom (vi), $\Phi\vdash\neg\bigwedge u.\neg(\varphi:u)$, where $u\in V_J\smallsetminus fvar_J(\varphi)$. Suppose $\overline{\varphi}\notin\bigcup_{c\in C}REASON_{\overline{c}}$. Then $\Phi\nvdash\varphi:c$ for all $c\in C$. Since $\Phi$ is a Henkin set, it follows that $\Phi\vdash\bigwedge u.\neg (\varphi:u)$. This contradiction shows that $\overline{\varphi}\in\bigcup_{c\in C}REASON_{\overline{c}}$.
From the axioms (xxi) and (xviii) and the fact that $\Phi$ is a Henkin set it easily follows that
\begin{equation*}
\overline{s}\le^\varLambda\overline{t}\Leftrightarrow REASON_{\overline{s}}\subseteq REASON_{\overline{t}}.
\end{equation*}
It remains to show that the map $\overline{s}\mapsto REASON_{\overline{s}}$ is injective. Suppose $\overline{s}\neq\overline{t}$, i.e. $\Phi\vdash\neg (s\equiv t)$. Then $\Phi\vdash\neg\forall x.(x:s\leftrightarrow x:t)$. Since $\Phi$ is a Henkin set, there is a propositional constant $d\in D$ such that $\Phi\vdash\neg (d:s\leftrightarrow d:t)$. That is, either ($\Phi\vdash d:s$ and $\Phi\nvdash d:t$) or ($\Phi\nvdash d:s$ and $\Phi\vdash d:t$). It follows that either $d\in REASON_{\overline{s}}\smallsetminus REASON_{\overline{t}}$ or $d\in REASON_{\overline{t}}\smallsetminus REASON_{\overline{s}}$. Thus, $REASON_{\overline{s}}\neq REASON_{\overline{t}}$. This finishes the proof of the Claim.\\
For an assignment $\gamma:V\rightarrow M\cup L$ let $\tau_\gamma:V\rightarrow Fm(C,D)\cup Tm(C)$ be a function with the property $\tau_\gamma(x)\in\gamma(x)$, for every $x\in V_P\cup V_J$. Notice that $\tau_\gamma:V_P\cup V_J\rightarrow M\cup L$ is both a substitution and an assignment. As the last ingredient of our model we define the Gamma-function by 
\begin{equation*}
\begin{split}
&\varGamma(\varphi,\gamma):=\overline{d},\text{ where }d\text{ is a propositional constant satisfying }d\approx_F \varphi[\tau_\gamma]\\
&\varGamma(s,\gamma):=\overline{c},\text{ where }c\text{ is a justification constant satisfying }c\approx_T s[\tau_\gamma].
\end{split}
\end{equation*}
It follows immediately that $\varGamma(\varphi,\gamma)=\overline{\varphi[\tau_\gamma]}$ and $\varGamma(s,\gamma)=\overline{s[\tau_\gamma]}$.\\
\textbf{Claim} 3: The Gamma-function satisfies the structure conditions of a model.\\
\textit{Proof of the Claim}: (HP) and (EP) follow immediately. In order to show (CP) let $\varphi\in Fm(C,D)$, $s\in Tm(C)$, and $\gamma$ and $\gamma'$ be assignments with $\gamma(x)=\gamma'(x)$ for all $x\in fvar(\varphi)$, and $\gamma(u)=\gamma'(u)$ for all $u\in var(s)$. Then $\tau_\gamma(x)\approx_F\tau_{\gamma'}(x)$ for all $x\in fvar_P(\varphi)$, $\tau_\gamma(v)\approx_T\tau_{\gamma'}(v)$ for all $v\in fvar_J(\varphi)$, and $\tau_\gamma(u)\approx_T\tau_{\gamma'}(u)$ for all $u\in var(s)$. That is, $\Phi\vdash\tau_{\gamma}\equiv_\varphi \tau_{\gamma'}$ and $\Phi\vdash\tau_{\gamma}\equiv_ s \tau_{\gamma'}$.  By the axioms (xii) and (xxii), $\varGamma(\varphi,\gamma)=\overline{\varphi[\tau_\gamma]}=\overline{\varphi[\tau_\gamma']}=\varGamma(\varphi,\gamma')$ and $\varGamma(s,\gamma)=\overline{s[\tau_\gamma]}= \overline{s[\tau_\gamma]}=\varGamma(s,\gamma')$. Now we aim for (SP). Let $\sigma:V\rightarrow Fm(C,D)\cup Tm(C)$ be a substitution, $\varphi\in Fm(C,D)$ and $s\in Tm(C)$. We must show: $\varGamma(\varphi[\sigma],\gamma)=\overline{\varphi[\sigma][\tau_\gamma]}=\overline{\varphi[\tau_{\gamma\sigma}]}=\varGamma(\varphi,\gamma\sigma)$ and $\varGamma(s[\sigma],\gamma)=\overline{s[\sigma][\tau_\gamma]}=\overline{s[\tau_{\gamma\sigma}]}=\varGamma(s,\gamma\sigma)$, for all assignments $\gamma :V\rightarrow M\cup L$. By Lemma \ref{240}, $\varphi[\sigma] [\tau_\gamma]=\varphi[\sigma\circ\tau_\gamma]$ and $s[\sigma] [\tau_\gamma]=s[\sigma\circ\tau_\gamma]$. Thus, it is enough to show that
\begin{equation*}
\begin{split}
&\varphi[\sigma\circ\tau_\gamma]\approx_F\varphi[\tau_{\gamma\sigma}]\text{ and }\\
&s[\sigma\circ\tau_\gamma]\approx_T s[\tau_{\gamma\sigma}].
\end{split}
\end{equation*}
Let $x\in fvar(\varphi)$. Then on the one hand $(\sigma\circ\tau_\gamma)(x)=\sigma(x)[\tau_\gamma]$. And on the other hand, $\tau_{\gamma\sigma}(x)\in\gamma\sigma(x)=\varGamma(\sigma(x),\gamma)=\overline{\sigma(x)[\tau_\gamma]}$. Hence, $\Phi\vdash(\sigma\circ\tau_\gamma)(x)\equiv\tau_{\gamma\sigma}(x)$, for each $x\in fvar(\varphi)$. Axiom (xii) yields the first assertion. The second assertion follows analogously using axiom (xii). Thus, (SP) holds. Let $\varphi,\psi$ be formulas such that $\varphi\prec\psi$. By Lemma \ref{320}, $\varphi[\tau_\gamma]\prec\psi[\tau_\gamma]$, where $\gamma$ is any assignment. By axiom (viii), $\varphi[\tau_\beta]<\psi[\tau_\beta]\in\Phi$. Thus, $\varGamma(\varphi,\gamma)=\overline{\varphi[\tau_\gamma]}<^\mathcal{M}\overline{\psi[\tau_\gamma]}=\varGamma(\psi,\gamma)$, and (RP) holds. \\
\textbf{Claim} 4: The Gamma-function satisfies the truth conditions.\\
\textit{Proof of the Claim}: 
\begin{equation*}
\begin{split}
\varGamma(\varphi\equiv\psi,\gamma)\in TRUE &\Leftrightarrow(\overline{\varphi\equiv\psi)[\tau_\gamma]}\in TRUE\\
&\Leftrightarrow\varphi[\tau_\gamma]\equiv\psi[\tau_\gamma]\in\Phi\\
&\Leftrightarrow\varphi[\tau_\gamma]\approx_F\psi[\tau_\gamma]\\
&\Leftrightarrow\varGamma(\varphi,\gamma)=\varGamma(\psi,\gamma).
\end{split}
\end{equation*}

\begin{equation*}
\begin{split}
\varGamma(\varphi <\psi,\gamma)\in TRUE &\Leftrightarrow\varphi[\tau_\gamma] <\psi[\tau_\gamma]\in\Phi\\
&\Leftrightarrow\overline{\varphi[\tau_\gamma]} <^\mathcal{M}\overline{\psi[\tau_\gamma]}\\
&\Leftrightarrow\varGamma(\varphi,\gamma) <^\mathcal{M}\varGamma(\psi,\gamma).
\end{split}
\end{equation*}

\begin{equation*}
\begin{split}
\varGamma(\square\varphi,\gamma)\in TRUE &\Leftrightarrow\square\varphi[\tau_\gamma]\in\Phi\\
&\Leftrightarrow\overline{\varphi[\tau_\gamma]}\in NECESSARY\\
&\Leftrightarrow\varGamma(\varphi,\gamma)\in NECESSARY.
\end{split}
\end{equation*}

\begin{equation*}
\begin{split}
\varGamma(\varphi:s,\gamma)\in TRUE &\Leftrightarrow (\varphi:s) [\tau_\gamma]\in\Phi\\
&\Leftrightarrow\varphi[\tau_\gamma]:s[\tau_\gamma]\in\Phi\\
&\Leftrightarrow\overline{\varphi[\tau_\gamma]}\in REASON_{\overline{s[\tau_\gamma]}}\\
&\Leftrightarrow\varGamma(\varphi,\gamma)\in REASON_{\varGamma(s,\gamma)}.
\end{split}
\end{equation*}

\begin{equation*}
\begin{split}
&\varGamma(\varphi\rightarrow\psi,\gamma)\in REASON_{\varGamma(s,\gamma)}\text{ and }\varGamma(\varphi,\gamma)\in REASON_{\varGamma(t,\gamma)}\\ &\Leftrightarrow\overline{\varphi[\tau_\gamma]\rightarrow\psi[\tau_\gamma]}\in REASON_{\overline{s [\tau_\gamma]}}\text{ and }\overline{\varphi[\tau_\gamma]}\in REASON_{\overline{t [\tau_\gamma]}} \\
&\Leftrightarrow (\varphi[\tau_\gamma]\rightarrow\psi[\tau_\gamma]) : s [\tau_\gamma]\in\Phi\text{ and }\varphi[\tau_\gamma] : t [\tau_\gamma]\in\Phi\\
&\Rightarrow\psi[\tau_\gamma] : (s[\tau_\gamma]\cdot t[\tau_\gamma])\in\Phi,\text{ by axiom (iii)}\\
&\Leftrightarrow (\psi : s\cdot t)[\tau_\gamma]\in\Phi\\
&\Leftrightarrow\overline{(\psi : s\cdot t) [\tau_\gamma]}\in TRUE\\
&\Leftrightarrow\varGamma(\psi:s\cdot t,\gamma)\in TRUE\\
&\Leftrightarrow\varGamma(\psi,\gamma)\in REASON_{\varGamma(s\cdot t,\gamma)}.
\end{split}
\end{equation*}

\begin{equation*}
\begin{split}
&\varGamma(\varphi,\gamma)\in REASON_{\varGamma(s,\gamma)}\cup REASON_{\varGamma(t,\gamma)}\\ 
&\Leftrightarrow\overline{\varphi[\tau_\gamma]}\in REASON_{s [\tau_\gamma]}\cup REASON_{t [\tau_\gamma]} \\
&\Leftrightarrow \varphi[\tau_\gamma] : s [\tau_\gamma]\in\Phi\text{ or }\varphi[\tau_\gamma] : t [\tau_\gamma]\in\Phi\\
&\Rightarrow \varphi[\tau_\gamma] : (s [\tau_\gamma] +  t [\tau_\gamma])\in\Phi,\text{ by axioms (iv), (v)}\\
&\Leftrightarrow(\varphi : s+t) [\tau_\gamma]\in\Phi\\
&\Leftrightarrow\overline{(\varphi : s+t) [\tau_\gamma]}\in TRUE\\
&\Leftrightarrow\varGamma(\varphi: s+t),\gamma)\in TRUE\\
&\Leftrightarrow\varGamma(\varphi,\gamma)\in REASON_{\varGamma(s + t,\gamma)}.
\end{split}
\end{equation*}

\begin{equation*}
\begin{split}
&\varGamma(\bigwedge u.\varphi,\gamma)\in TRUE\\
&\Leftrightarrow (\bigwedge u.\varphi) [\tau_\gamma]\in\Phi\\
&\Leftrightarrow \bigwedge v.(\varphi[\tau_\gamma[u:=v]])\in\Phi\\
&\overset{(*)}{\Leftrightarrow}\varphi[\tau_\gamma[u:=v]][v:=c]\in\Phi,\text{ for all }c\in C,\text{ since }\Phi\text{ is a Henkin set}\\
&\Leftrightarrow\varphi[\tau_\gamma[u:=c]]\in\Phi,\text{ for all }c\in C\\
&\overset{(**)}{\Leftrightarrow}\varphi[\tau_{\gamma_u^{\overline{c}}}]\in\Phi,\text{ for all }c\in C\\
&\Leftrightarrow\varGamma(\varphi,\gamma_u^{\overline{c}})\in TRUE,\text{ for all }\overline{c}\in L
\end{split}
\end{equation*}

It remains to show that the equivalences (*) and (**) hold.\\
(*): $v$ is the variable forced by the substitution $\tau_\gamma$ w.r.t. $\bigwedge u.\varphi$. Thus, $v\notin fvar_J(\varphi[\tau_\gamma])$. Then it is clear that the equivalence (*) holds.\\ 
(**): Let $z\in fvar(\varphi)$. First, we suppose $z\neq u$. Then $\tau_\gamma[u:=c](z)=\tau_\gamma(z)\in\gamma(z)$ and $\tau_{\gamma_u^{\overline{c}}}(u)\in\gamma_u^{\overline{c}}(u)=\gamma(u)$. Thus, $\tau_\gamma[u:=c](z)\approx_F\tau_{\gamma_x^{\overline{c}}}(z)$ if $z\in fvar_P(\varphi)$; and $\tau_\gamma[u:=c](z)\approx_T\tau_{\gamma_x^{\overline{c}}}(z)$ if $z\in fvar_J(\varphi)$. Now suppose $z=u$. Then $\tau_\gamma[u:=c](z)=c$ and $\tau_{\gamma_u^{\overline{c}}}(z)\in\gamma_u^{\overline{c}}(z)=\overline{c}$. Again, $\tau_\gamma[x:=c](z)\approx_T\tau_{\gamma_x^{\overline{c}}}(z)$. By axiom (xii), $\varphi[\tau_\gamma[u:=c]]\approx_F\varphi[\tau_{\gamma_u^{\overline{c}}}]$. Item (v) of Lemma \ref{600} yields the equivalence (**).

The condition concerning the propositional quantifier follows analogously.  Finally,
\begin{equation*}
\varGamma(s\le t,\gamma)\in TRUE\Leftrightarrow s [\tau_\gamma] \le t [\tau_\gamma] \in\Phi\Leftrightarrow\varGamma(s,\gamma)\le^\varLambda\varGamma(t,\gamma).
\end{equation*}
The remaining truth conditions follow easily from axioms of propositional logic. Consider now the specific assignment $\beta:V\rightarrow M\cup L$ defined by
\begin{equation*}
\beta(x):=\overline{x}
\end{equation*}
for each $x\in V_P\cup V_J$.\\
\textbf{Claim} 5: $\varphi[\tau_\beta]\approx_F\varphi$, for all $\varphi\in Fm(C,D)$.\\
\textit{Proof of the Claim}: From the definition of the assignment/substitution $\tau_\beta$ it follows that $\Phi\vdash\tau_\beta(x)\equiv \varepsilon(x)$ for all $x\in fvar(\varphi)$, where $\varepsilon$ is the identity substitution. The Claim now follows from axiom (xii).\\
We have shown that 
\begin{equation*}
\mathcal{M}:=(\varLambda,M,TRUE,NECESSARY,(REASON_l)_{l\in L},FALSE, <^\mathcal{M}, \varGamma)
\end{equation*}
is a model. It follows now from Claim 5 and item (v) of Lemma \ref{600} that the interpretation $(\mathcal{M},\beta)$ is a model of the Henkin set $\Phi$:
\begin{equation*}
(\mathcal{M},\beta)\vDash\varphi\Leftrightarrow\varGamma(\varphi,\beta)=\overline{\varphi[\tau_\beta]}\in TRUE\Leftrightarrow\varphi[\tau_\beta]\in\Phi\Leftrightarrow\varphi\in\Phi.
\end{equation*}

\begin{theorem}\label{640}
Every consistent set has a model.
\end{theorem}

\paragraph*{Proof.}
Let $\Phi\subseteq Fm(C,D)$ be consistent. We show that $\Phi$ extends to a Henkin set $\Phi^*$ in an appropriate extended language $Fm(C^*,D^*)$. By Theorem \ref{620}, $\Phi^*$ has a model. We will see that its
\textit{reduct} to the sublanguage $Fm(C,D)$ is a model of $\Phi$. \\ 
Let $C_0:=C$, $D_0:=D$, $\Phi_0:=\Phi$. If $C_n$, $D_n$ and $\Phi_n\subseteq Fm(C_n,D_n)$ are already defined, then define 
\begin{equation*}
\begin{split}
&C_{n+1}:=C_n\cup\{c_{\varphi,u}\mid\varphi\in Fm(C_n,D_n), u\in fvar_J(\varphi)\}\\
&D_{n+1}:=D_n\cup\{d_{\varphi,d}\mid\varphi\in Fm(C_n,D_n), d\in fvar_P(\varphi)\}\\
&\Phi_{n+1}:=\Phi_n\cup X(C_n,D_n)\cup Y(C_n,D_n)
\end{split}
\end{equation*}
according to the notation of Definition \ref{540}. By Lemma \ref{560}, $\Phi_n\cup X(C_n,D_n)\subseteq Fm(C_{n+1},D_n)$ and $(\Phi_n\cup X(C_n,D_n))\cup Y(C_n,D_n)\subseteq Fm(C_{n+1},D_{n+1})$ are consistent. That is, $\Phi_{n+1}$ is consistent in $Fm(C_{n+1},D_{n+1})$. Finally, we put $\Phi^+:=\bigcup_{n<\omega}\Phi_n$. It follows that $\Phi^+\subseteq Fm(C^*,D^*)$, where $C^*=\bigcup_{n<\omega}C_n$, $D^*=\bigcup_{n<\omega} D_n$. Since derivation is finitary, $\Phi^+$ is consistent in the language $Fm(C^*,D^*)$. By a standard argument that uses Zorn's Lemma, $\Phi^+$ extends to a maximally consistent set $\Phi^*\subseteq Fm(C^*,D^*)$. 

If $\Phi^*\vdash\bigwedge u.\varphi$, then by axiom (xiv), $\Phi^*\vdash\varphi[u:=c]$, for all $c\in C^*$. The other way around, suppose $\Phi^*\vdash\varphi[u:=c]$ for all $c\in C^*$, where $u\in fvar_J(\varphi)$. Let $n, m$ be minimal with the property $\varphi\in Fm(C_n,D_m)$. If $n\ge m$, then $\varphi\in Fm(C_n, D_n)$, $\varphi[u:=c_{\varphi,u}]\in Fm(C_{n+1},D_n)$ and $c_{\varphi,u}\in C_{n+1}\smallsetminus C_n$. By construction, $\neg (\varphi^u)\in X(C_n,D_n)\subseteq\Phi_{n+1}\subseteq\Phi^*$. Thus, $\Phi^*\vdash\neg(\varphi^u)$. If $n<m$, then $\varphi\in Fm(C_m, D_m)$, $\varphi[u:=c_{\varphi,u}]\in Fm(C_{m+1},D_m)$ and $c_{\varphi,u}\in C_{m+1}\smallsetminus C_m$. Again, it follows that $\Phi^*\vdash\neg(\varphi^u)$. Towards a contradiction suppose $\Phi^*\nvdash\bigwedge u.\varphi$. Since $\Phi^*$ is maximally consistent, $\Phi^*\vdash\neg\bigwedge u.\varphi$. Since $\Phi^*\vdash\varphi[u:=c]$ for all $c\in C^*$, we have in particular $\Phi^*\vdash\varphi[u:=c_{\varphi,u}]$. Thus, $\Phi^*\vdash\neg\bigwedge u.\varphi\wedge\varphi[u:=c_{\varphi,u}]$. That is, $\Phi^*\vdash\varphi^u$. This is a contradiction to $\Phi^*\vdash\neg(\varphi^u)$ and the consistency of $\Phi^*$. Therefore, $\Phi^*\vdash\bigwedge u.\varphi$. We have shown that $\Phi^*$ has the properties of a Henkin set. 

Let $(\mathcal{M}^*,\beta)$ be a model of $\Phi^*$ w.r.t. the language $Fm(C^*,D^*)$, and let $\varGamma^*$ be the Gamma-function of $\mathcal{M}^*$. Let $\varGamma$ be the restriction of $\varGamma^*$ to the domain $Fm(C,D)\subseteq Fm(C^*,D^*)$. If we replace $\varGamma^*$ with $\varGamma$ in $\mathcal{M}^*$, then obviously we get a model $\mathcal{M}$ such that $(\mathcal{M},\beta)\vDash\varphi\Leftrightarrow(\mathcal{M}^*,\beta)\vDash\varphi$ for all formulas $\varphi\in Fm(C,D)$. The model $\mathcal{M}$ is called the reduct of $\mathcal{M}^*$ to the sublanguage $Fm(C,D)$. In particular, $(\mathcal{M},\beta)\vDash\Phi$. $\square$

\begin{theorem}[Completeness Theorem]\label{660}
For all $\Phi\cup\{\varphi\}\subseteq Fm(C,D)$:
\begin{equation*}
\Phi\Vdash\varphi\Leftrightarrow\Phi\vdash\varphi.
\end{equation*}
\end{theorem}

\paragraph*{Proof.}
The direction from right to left follows from the fact that all axioms are valid and that the rule of Modus Ponens is sound. Let $\Phi\Vdash\varphi$. Suppose $\Phi\nvdash\varphi$. From axioms of classical propositional logic it follows that $\Phi\cup\{\neg\varphi\}$ is consistent. By the preceding results, this set has a model. This is a contradiction to $\Phi\Vdash\varphi$. $\square$

\section{Capturing the modal logics S4 and S5}

In this last section we discuss some extensions of our original deductive system. In particular, we will show that adding the following axiom schema (4) and the rule of Axiom Necessitation (see below) results in a system that is able to restore modal logic S4. The condition that formulas of the form (4) are theorems seems to be essential for capturing modal logics by our semantics. We are unable to restore weaker modal logics such as $T$ or $K$ not containing (4). Axiom schema (4) is given by all formulas of the form
\begin{equation*}
\square\varphi\rightarrow\square\square\varphi
\end{equation*}
In standard modal logic this schema stands for $S_4$. We denote the system that we get by adding (4) to our system $Ax$ (together with Modus Ponens) by $Ax +(4)$. Let $\vdash_{(4)}$ be the resulting relation of derivability. The new axiom schema corresponds to the following new truth condition (4) of a model $\mathcal{M}$:
\begin{equation*}
\varGamma(\square\varphi,\gamma)\in TRUE\Rightarrow\varGamma(\square\square\varphi,\gamma)\in TRUE,
\end{equation*} 
for all $\varphi\in Fm(C,D)$ and for all assignments $\gamma:V\rightarrow M$. Let $\Vdash_{(4)}$ be the consequence relation of the logic generated by all models that satisfy the additional truth condition $(4)$.

\begin{corollary}\label{680}
For all $\Phi\cup\{\varphi\}\subseteq Fm(C,D)$:
\begin{equation*}
\Phi\Vdash_{(4)}\varphi\Leftrightarrow\Phi\vdash_{(4)}\varphi.
\end{equation*}
\end{corollary}

\paragraph*{Proof.}
One easily checks that $\square\varphi\rightarrow\square\square\varphi$ is valid in all models with truth condition $(4)$. Towards completeness we follow exactly the same strategy as above. It is sufficient to show that the model of $\Phi$ constructed in Theorem \ref{620} (where $\Phi$ is now a Henkin set w.r.t. the system $Ax+(4)$) satisifies the new truth condition (4):
\begin{equation*}
\begin{split}
\varGamma(\square\varphi,\gamma)\in TRUE &\Leftrightarrow\overline{\square\varphi[\tau_\gamma]}\in TRUE\\
&\Leftrightarrow\square\varphi[\tau_\gamma]\in\Phi\\
&\Rightarrow\square\square\varphi[\tau_\gamma]\in\Phi,\text{ by the new axiom }\square\varphi\rightarrow\square\square\varphi\\
&\Leftrightarrow\overline{\square\square\varphi[\tau_\gamma]}\in TRUE\\
&\Leftrightarrow\varGamma(\square\square\varphi,\gamma)\in TRUE.
\end{split}
\end{equation*}
$\square$

Recall that the following axiom schema (E) stands for the modal logic $S_5$: $\diamondsuit\varphi\rightarrow\square\diamondsuit\varphi$. Adding all formulas of this form as axioms to our system $Ax$ results in a system that we denote by $Ax+(E)$. Axiom schema (E) corresponds to the following new truth condition (E) of a model $\mathcal{M}$:
\begin{equation*}
\varGamma(\varphi,\gamma)\in POSSIBLE\Rightarrow\varGamma(\diamondsuit\varphi,\gamma)\in NECESSARY,
\end{equation*} 
for all $\varphi\in Fm(C,D)$ and for all assignments $\gamma:V\rightarrow M$. Let $\vdash_{(E)}$ be the derivability relation of system $Ax+(E)$, and let $\Vdash_{(E)}$ be the consequence relation of the logic generated by all models that satisfy the truth condition (E). Then in a similar way as above one proves   

\begin{corollary}\label{700}
For all $\Phi\cup\{\varphi\}\subseteq Fm(C,D)$:
\begin{equation*}
\Phi\Vdash_{(E)}\varphi\Leftrightarrow\Phi\vdash_{(E)}\varphi.
\end{equation*}
\end{corollary}

Axiom $K$ of modal logic, $\square(\varphi\rightarrow\psi)\rightarrow(\square\varphi\rightarrow\square\psi)$, is not an axiom of our systems but it is a valid (i.e., true in all models) and therefore derivable from $Ax$, i.e. a theorem.
This is shown in detail in the proof of Theorem \ref{820} below. A further essential ingredient of current modal logics is the rule of Necessitation: from $\varphi$ derive $\square\varphi$. In the following, we will show that the Necessitation rule can be restored by introducing a rule of Axiom Necessitation together with axiom schema (4). 

\begin{definition}\label{720}
Let $\Phi\subseteq Fm(C,D)$. $\Phi^{\vdash^{(AxNec)}_{(4)}}$ is the smallest set containing $Ax\cup\Phi + (4)$ and being closed under Modus Ponens and under the rule of Axiom Necessitation: If $\chi\in Ax + (4)$, then $\square\chi\in\Phi^{\vdash^{(AxNec)}_{(4)}}$. Instead of $\varphi\in\Phi^{\vdash^{(AxNec)}_{(4)}}$ we write $\Phi\vdash^{(AxNec)}_{(4)}\varphi$.
\end{definition}

The semantic counterpart is given by the class of all models with truth condition (4) and the following truth condition (AxNec):\\ 
If $\chi\in Ax + (4)$, then $\varGamma(\square\chi,\gamma)\in TRUE$, for all assignments $\gamma:V\rightarrow M$.

Truth condition (AxNec) guarantees the soundness of the rule of Axiom Necessitation (AxNec). On the other hand, following the proof of Theorem \ref{620}, one recognizes that the model constructed there satisfies truth condition (AxNec), given that the rule of Axiom Necessitation is part of the deductive system. Let $\Vdash^{(AxNec)}_{(4)}$ be the consequence relation of the logic generated by the class of all models satisfying truth conditions (4) and (AxNec).

\begin{corollary}\label{740}
For all $\Phi\cup\{\varphi\}\subseteq Fm(C,D)$:
\begin{equation*}
\Phi\Vdash^{(AxNec)}_{(4)}\varphi\Leftrightarrow\Phi\vdash^{(AxNec)}_{(4)}\varphi.
\end{equation*}
\end{corollary}

In the same way, we define $\vdash^{(AxNec)}_{(E)}$ and $\Vdash^{(AxNec)}_{(E)}$, and obtain a corresponding completeness result. We denote the respective deductive systems by $Ax + (AxNec) + (4)$ and $Ax + (AxNec) + (E)$.  

The system $Ax + (AxNec) + (4)$ involves the following fundamental principle of modal logic.

\begin{theorem}[Necessitation]\label{760}
For every $\varphi\in Fm(C,D)$:
\begin{equation*}
\vdash_{(4)}^{(AxNec)}\varphi \Longrightarrow\text{ }\vdash_{(4)}^{(AxNec)}\square\varphi.
\end{equation*}
\end{theorem}

\paragraph*{Proof.}
Suppose $\vdash_{(4)}^{(AxNec)}\varphi$. We show the assertion by induction on the length $n$ of the derivation of $\varphi$. If $n=1$, then $\varphi$ is an axiom or $\varphi$ is derived by the rule (AxNex). In the former case, (AxNec) yields $\vdash_{(4)}^{(AxNec)}\square\varphi$. In the latter case, $\varphi=\square\psi$ for some axiom $\psi$. By axiom (4), $\square\psi\rightarrow\square\square\psi$. Modus Ponens yields $\vdash_{(4)}^{(AxNec)}\square\square\psi$, that is, $\vdash_{(4)}^{(AxNec)}\square\varphi$.

Now suppose there are formulas $\psi$ and $\psi\rightarrow\varphi$, derived in at most $n\ge 1$ steps, and $\varphi$ is obtained by Modus Ponens. By induction hypothesis, $\square\psi$ and $\square(\psi\rightarrow\varphi)$ are derivable from the empty set. The formula $\square(\psi\rightarrow\varphi)\rightarrow(\square\psi\rightarrow\square\varphi)$ is a theorem (see the proof of Theorem \ref{820} below). Modus Ponens yields $\vdash_{(4)}^{(AxNec)}\square\varphi$. $\square$\\

The last result implies that adding axiom (4) and the rule of Axiom Necessitation to our original system we can derive in particular all theorems of modal logic S4. Recall that axiom schema (4) is derivable in modal logic $S_5$ (in fact, S4 is contained in S5). Thus, from the proof of Theorem \ref{760} it follows that we are also able to capture modal logic $S_5$. Taking into account our soundness and completeness results we conclude the following.

\begin{corollary}\label{780}
Let $\mathcal{L}\subseteq Fm(C,D)$ be the language of propositional modal logic with the set of propositional variables $V_P$. Then for all $\varphi\in\mathcal{L}$, 
\begin{equation*}
\begin{split}
\varphi\text{ is a theorem of }S_4&\Longleftrightarrow\text{ }\vdash_{(4)}^{(AxNec)}\varphi \Longleftrightarrow\text{ }\Vdash_{(4)}^{(AxNec)}\varphi\\
\varphi\text{ is a theorem of }S_5&\Longleftrightarrow\text{ }\vdash_{(E)}^{(AxNec)}\varphi \Longleftrightarrow\text{ }\Vdash_{(E)}^{(AxNec)}\varphi
\end{split}
\end{equation*}
\end{corollary}

In the rest of this section we present an alternative, purely model-theoretic proof of this fact restricting our attention to modal logic S4. More precisely, we are going to construct for each model of modal logic S4 an interpretation $(\mathcal{M},\beta)$ of our non-Fregean logic with truth conditions (4) and (AxNec) satisfying the same set of formulas of the language $\mathcal{L}$ of modal logic, and vice-versa. This will give us some interesting insights into the connection between the non-Fregean semantics of our $\in_J$-Logic and the possible worlds semantics of modal logic S4.
As above, let $\mathcal{L}$ be the language of propositional modal logic, where $V_P=\{x_0,x_1,...\}$ is the set of propositional variables. Recall that a frame of modal logic S4 is a structure $\mathcal{F}=(W,R)$, where $W$ is a set of worlds and $R\subseteq W\times W$ is a reflexive and transitive accessible relation on $W$. A truth value assignment of a frame $\mathcal{F}=(W,R)$ is a function $g:W\rightarrow(V_P\rightarrow\{0,1\})$, $w\mapsto g_w\in 2^{V_P}$. The satisfaction relation $(w,g)\Vdash\varphi$ is inductively defined as follows: $(w,g)\Vdash x :\Leftrightarrow g_w(x)=1$, $(w,g)\Vdash\neg\varphi :\Leftrightarrow (w,g)\nVdash\varphi$, $(w,g)\Vdash\varphi\rightarrow\psi :\Leftrightarrow (w,g)\nVdash\varphi$ or $(w,g)\Vdash\psi$, $(w,g)\Vdash\square\varphi :\Leftrightarrow (w',g)\Vdash\varphi$ for all $w'\in W$ with $wRw'$.

\begin{theorem}\label{800}
Let $w$ be a world of a given frame $\mathcal{F}=(W,R)$ of modal logic S4, and let $g:W\rightarrow (V_P\rightarrow\{0,1\})$ be a truth value assignment. There exists a model $\mathcal{M}$ with the additional truth conditions (4) and (AxNec), and an assignment $\beta:V_P\rightarrow M$ such that for all $\varphi\in\mathcal{L}$:
\begin{equation*}
(\mathcal{M},\beta)\vDash\varphi\Leftrightarrow (w,g)\Vdash\varphi.
\end{equation*}
\end{theorem}

\paragraph*{Proof.}
Let $t, f, nec, imp$ be (names of) propositions, where $t$ stands for \textit{true}, $f$ stands for \textit{false}, $nec$ stands for \textit{necessary}, and $imp$ stands for \textit{impossible}. Put $TRUE:=\{t,nec\}$, $FALSE:=\{f,imp\}$ and $NECESSARY:=\{nec\}$. The propositional universe of our model is $M:=TRUE\cup FALSE$ and the reference relation is given by $<^\mathcal{M}:=M\times M$. Furthermore, we define $L:=\{l\}$, $l\le^{\varLambda} l$, $l+^{\varLambda} l:=l$,  $l\cdot^{\varLambda} l:=l$, and $REASON_l:=\{nec\}$, where $l$ is any new symbol that will serve as a name for the (unique) justification $\{nec\}$. We define the Gamma-function inductively on the construction of formulas, simultaneously for all assignments $\gamma:V\rightarrow M$:

\begin{equation*}
\begin{split}
\varGamma(x,\gamma) &:=\gamma(x),\text{ for }x\in V_P\cup V_J\\
\varGamma(c)&:=l,\text{ for }c\in C\\
\varGamma(d)&:=t,\text{ for }d\in D \text{  (this assignment is arbitrary)}\\
\varGamma(\varphi:true,\gamma)&:=\varGamma(\varphi,\gamma)\\
\varGamma(\varphi:false,\gamma)&:=\varGamma(\neg\varphi,\gamma):=
\begin{cases}
\begin{split}
&t,\text{ if }\varGamma(\varphi,\gamma)=f\\
&f,\text{ if }\varGamma(\varphi,\gamma)=t\\
&nec,\text{ if }\varGamma(\varphi,\gamma)=imp\\
&imp,\text{ if }\varGamma(\varphi,\gamma)=nec
\end{split}
\end{cases}
\end{split}
\end{equation*}

\begin{equation*}
\begin{split}
\varGamma(\varphi\rightarrow\psi,\gamma)&:=
\begin{cases}
\begin{split}
&t,\text{ if }\varGamma(\varphi,\gamma)=f\text{ or }\varGamma(\psi,\gamma)=t\\
&f,\text{ if }\varGamma(\varphi,\gamma)=t\text{ and }\varGamma(\varphi,\gamma)=f\\
&nec,\text{ if }\varGamma(\varphi,\gamma)=imp\text{ or }\varGamma(\psi,\gamma)=nec\\
&imp,\text{ if }\varGamma(\varphi,\gamma)=nec\text{ and }\varGamma(\psi,\gamma)=imp
\end{split}
\end{cases}
\end{split}
\end{equation*}

\begin{equation*}
\begin{split}
\varGamma(\varphi\equiv\psi,\gamma)&:=
\begin{cases}
\begin{split}
&t,\text{ if }\varGamma(\varphi,\gamma)=\varGamma(\psi,\gamma)\\
&f,\text{ else }
\end{split}
\end{cases}
\end{split}
\end{equation*}

\begin{equation*}
\varGamma(\varphi <\psi,\gamma):=\varGamma(s\le t,\gamma):=t
\end{equation*}

\begin{equation*}
\begin{split}
\varGamma(\square\varphi,\gamma)&:=\varGamma(\varphi:s,\gamma):=
\begin{cases}
\begin{split}
&nec,\text{ if }\varGamma(\varphi,\gamma)=nec\text{ or }\varphi\in Ax + (4)\\
&f,\text{ else }
\end{split}
\end{cases}
\end{split}
\end{equation*}

\begin{equation*}
\begin{split}
\varGamma(\bigwedge u.\varphi,\gamma):=
\begin{cases}
\begin{split}
&t,\text{ if }\varGamma(\varphi,\gamma_u^l)\in TRUE\\
&f,\text{ else }
\end{split}
\end{cases}
\end{split}
\end{equation*}

\begin{equation*}
\begin{split}
\varGamma(\forall x.\varphi,\gamma)&:=
\begin{cases}
\begin{split}
&t,\text{ if }\varGamma(\varphi,\gamma_x^m)\in TRUE\text{ for all }m\in M\\
&f,\text{ else }
\end{split}
\end{cases}
\end{split}
\end{equation*}

It is crucial that the axiom $\square\varphi\rightarrow\square\square\varphi$ holds in the given model $w$. Otherwise, we were not able to define the Gamma-function in any reasonable way for the case $\varGamma(\square\varphi,\gamma)$. One easily checks that
\begin{equation*}
\mathcal{M}=(\varLambda,M,TRUE,NECESSARY,REASON_l,FALSE,<^\mathcal{M},\varGamma)
\end{equation*}
satisfies all properties of a model. The structure conditions (CP) and (SP) follow by induction. Only the proof of the quantifier case of (SP) is somewhat complicated: suppose $\varphi=\forall x.\psi$, and let $\sigma:V\rightarrow M\cup L$ be a substitution. \\
\textbf{Claim}: For all $m\in M$ and all $y\in fvar(\psi)$: 
\begin{equation}\label{100}
(\gamma\sigma)_x^m(y)=(\gamma_z^m\sigma[x:=z])(y),
\end{equation}
where $z$ is the variable forced by $\sigma$ w.r.t. $\varphi$. That is, $\varphi[\sigma]=(\forall x.\psi)[\sigma]=\forall z.(\psi[\sigma[x:=z])$. Let $y\in fvar(\psi)$. First, suppose $y=x$. Then $(\gamma\sigma)_x^m(y)=m$. On the other hand, $(\gamma_z^m\sigma[x:=z])(y)=\varGamma(\sigma[x:=z](y),\gamma_z^m)=\varGamma(z,\gamma_z^m)=\gamma_z^m(z)=m$. Now suppose that $y\neq x$. Note that by definition, $z\notin fvar(\sigma(y))$. Then by (CP) we get $(\gamma\sigma)_x^m(y)=(\gamma\sigma)(y)=\varGamma(\sigma(y),\gamma)=\varGamma(\sigma(y), \gamma_z^m)=\varGamma(\sigma[x:=z](y),\gamma_z^m)=(\gamma_z^m\sigma[x:=z])(y)$. This proves the Claim. \\
Consequently:
\begin{equation*}
\begin{split}
\varGamma(\forall x.\psi,\gamma\sigma)=t &\Leftrightarrow\varGamma(\psi,(\gamma\sigma)_x^m)=t,\text{ for all } m\in M\\ 
&\Leftrightarrow\varGamma(\psi,\gamma_z^m\sigma[x:=z])=t,\text{ for all } m\in M, \text{ by \eqref{100} and (CP)}\\
&\Leftrightarrow\varGamma(\psi[\sigma[x:=z]],\gamma_z^m)=t, \text{ by induction hypothesis}\\
&\Leftrightarrow\varGamma(\forall z.\psi[\sigma[x:=z]],\gamma_z^m)=t\\
&\Leftrightarrow\varGamma((\forall x.\psi)[\sigma],\gamma)=t
\end{split}
\end{equation*}
Since there are only the two possible truth values $t$ or $f$ for a quantified formula, it follows that $\varGamma(\varphi[\sigma],\gamma)=\varGamma(\varphi,\gamma\sigma)$. The case $\varphi=\bigwedge u.\psi$ follows similarly. Thus, (SP) holds. The remaining structure properties of a model are trivially satisfied. The truth conditions follow readily from the construction. From the definition of the Gamma-function in the case $\varGamma(\square\varphi,\gamma)$ it follows that the model also satisfies the truth conditions (4) and (AxNec).

Now we define the assignment $\beta:V_P\cup V_J\rightarrow M\cup L$ by
\begin{equation*}
\begin{split}
&\beta(x):=
\begin{cases}
\begin{split}
&t,\text{ if }(w,g)\Vdash x\text{ and }(w,g)\nVdash\square x\\
&f,\text{ if }(w,g)\nVdash x\text{ and }(w,g)\nVdash\square\neg x\\
&nec,\text{ if }(w,g)\Vdash\square x\\
&imp,\text{ if }(w,g)\Vdash\square\neg x
\end{split}
\end{cases}
\end{split}
\end{equation*}

Then one shows by induction on $\varphi\in\mathcal{L}$ that the following holds:
\begin{equation*}
\begin{split}
&\varGamma(\varphi,\beta)=t\Leftrightarrow (w,g)\Vdash\varphi\text{ and }(w,g)\nVdash\square \varphi\\
&\varGamma(\varphi,\beta)=f\Leftrightarrow (w,g)\nVdash\varphi\text{ and }(w,g)\nVdash\square\neg\varphi\\
&\varGamma(\varphi,\beta)=nec\Leftrightarrow (w,g)\Vdash\square\varphi\\
&\varGamma(\varphi,\beta)=imp\Leftrightarrow (w,g)\Vdash\square\neg\varphi
\end{split}
\end{equation*}

This implies

\begin{equation*}
(\mathcal{M},\beta)\vDash\varphi\Leftrightarrow\varGamma(\varphi,\beta)\in TRUE\Leftrightarrow (w,g)\Vdash\varphi,
\end{equation*}
for every $\varphi\in \mathcal{L}$. $\square$\\

It might be an interesting observation that the model constructed in the proof of Theorem \ref{800} is in some sense a $4$-valued model of a logic that extends the modal system S4.  

Now we prove the converse of Theorem \ref{800}. As above, we suppose that $\mathcal{L}$ is the language of propositional modal logic with $V_P=\{x_0, x_1, ... \}$ the set of propositional variables.

\begin{theorem}\label{820}
Let $\mathcal{M}$ be a model satisfying the additional truth conditions (AxNec) and (4). Let $\beta:V_P\rightarrow M$ be an assignment. Then there exists a frame $(W,R)$ of modal logic S4, a truth value assignment $g:W\rightarrow (V_P\rightarrow\{0,1\})$, and a world $w\in W$ such that for all $\varphi\in \mathcal{L}$: 
\begin{equation*}
(\mathcal{M},\beta)\vDash\varphi\Leftrightarrow (w,g)\Vdash\varphi.
\end{equation*}
\end{theorem}

\paragraph*{Proof.}
We will use some basic concepts of the theory of abstract logics (see, e.g., \cite{lewbru, lewjlc}). A classical abstract logic $\mathcal{A}=(Expr,Th,\{\rightarrowtail,\sim.\curlyvee,\curlywedge\})$ is given by a set of expressions or formulas $Expr$, a subset $Th$ of the powerset of $Expr$ such that for every non-empty $\mathcal{T}\subseteq Th$, $\bigcap\mathcal{T}\in Th$, and a set of connectives $\{\curlyvee,\curlywedge,\rightarrowtail,\sim\}$. The elements of $Th$ are called theories. Every theory is the intersection of a non-empty set of maximal theories (maximal w.r.t. set-theoretic inclusion). The connectives satisfy the following properties. For all expressions $a,b$ and all maximal theories $T$: $a\rightarrow b\in T\Leftrightarrow a\notin T$ or $b\in T$, $\sim a\in T\Leftrightarrow a\notin T$, $a\curlyvee b\in T\Leftrightarrow a\in T$ or $b\in T$, $a\curlywedge b\in T\Leftrightarrow a\in T$ and $b\in T$. Note that $Th\cup\{Expr\}$ is a closure system. The corresponding closure operator is the consequence relation of logic $\mathcal{A}$ which is required to be compact. A set $B$ of expressions is said to be consistent in $\mathcal{A}$ if $B$ is contained in some theory. It follows that a set $T$ of expressions is a theory iff $T$ is consistent and closed under the consequence relation. Note that $Expr$ is not a theory.

Obviously, the class of all models of $\in_J$-Logic generates a classical abstract logic $\mathcal{A}$. In fact, the set of formulas satisfied by a model is a maximal theory. The compactness of the consequence relation follows from our completeness theorem. The set of formulas $F_M:=\{\varphi\in Fm(C,D)\mid\varGamma(\varphi,\beta)\in TRUE\}=\{\varphi\in Fm(C,D)\mid (\mathcal{M},\beta)\vDash\varphi\}$ is a maximal theory, and $F_N:=\{\varphi\in Fm(C,D)\mid\varGamma(\varphi,\beta)\in NECESSARY\}\subseteq F_M$ is therefore a consistent set in the sense of the abstract logic $\mathcal{A}$. \\
\textbf{Claim} 1: $F_N$ is a theory in the sense of the classical abstract logic $\mathcal{A}$, i.e. $F_N$ is the intersection of a non-empty set of maximal theories.\\
\textit{Proof of the Claim}: Since a set $B$ is a theory iff $B$ is consistent and deductively closed, it remains to show that $F_N$ is deductively closed, i.e. $F_N\Vdash\varphi$ implies $\varphi\in F_N$. Recall that $\Vdash$ coincides with the closure operator of the closure system associated with the classical abstract logic $\mathcal{A}$. By our completeness theorem, it is enough to prove: $F_N\vdash\varphi$ implies $\varphi\in F_N$. We show this by induction on the length of a derivation. If this length is $1$, then $\varphi\in F_N$ or $\varphi$ is an axiom. We may assume that $\varphi$ is an axiom. Truth condition (AxNec) implies $\varphi\in F_N$. If the length of the derivation is greater than $1$, then there are formulas $\psi$ and $\psi\rightarrow\varphi$ such that $N_F\vdash\psi$ and $N_F\vdash\psi\rightarrow\varphi$. By induction hypothesis, $\psi,\psi\rightarrow\varphi\in F_N$. That is, $\varGamma(\psi,\beta)\in NECESSARY$ and $\varGamma(\psi\rightarrow\varphi,\beta)\in NECESSARY$. There are indexes $l,k$ such that $\varGamma(\psi\rightarrow\varphi,\beta)\in REASON_k$ and $\varGamma(\psi,\beta)\in REASON_l$. Let $u,v\in V_J\smallsetminus (fvar_J(\varphi)\cup fvar_J(\psi))$, and let $\beta'$ be an assignment that coincides with $\beta$ on $fvar(\varphi)\cup fvar(\psi)$, and $\beta'(u)=k$ and $\beta'(v)=l$. Then by (CP), $\varGamma(\psi\rightarrow\varphi,\beta')=\varGamma(\psi\rightarrow\varphi,\beta)\in REASON_{\varGamma(u,\beta')}=REASON_k$ and $\varGamma(\psi,\beta')=\varGamma(\psi,\beta)\in REASON_{\varGamma(v,\beta')}=REASON_l$. By truth condition (xi) and (HP) of a model, $\varGamma(\varphi,\beta)=\varGamma(\varphi,\beta')\in REASON_{\varGamma(u\cdot v,\beta')}=REASON_{\varGamma(u,\beta')\cdot^\Lambda\varGamma(v,\beta')}=REASON_{k\cdot^\Lambda l}\subseteq NECESSARY$. Then $\varphi\in F_N$ and $F_N$ is deductively closed. This proves the Claim.\footnote{Notice that we have in particular proved that $F_N$ is closed under Modus Ponens. This immediately implies that axiom $K$, $\square(\psi\rightarrow\varphi)\rightarrow(\square\psi\rightarrow\square\varphi)$, is valid and therefore a theorem in $\in_J$-Logic. We have already used this fact before.}\\

Since $F_N$ is a theory, there are maximal theories $(T_i)_{i\in I}$, $T_i\subseteq Fm(C,D)$, such that $F_N=\bigcap_{i\in I} T_i$. 
Observe that $(\mathcal{M},\beta)\vDash\square\varphi\Leftrightarrow\varphi\in F_N\Leftrightarrow\varphi\in T_i$ for all $i\in I$. For each $i\in I$ we define an extensional model $\mathcal{M}_i$ as follows. Let $t,f,l$ be new symbols. We define $M_i:=\{t,f\}$, $L_i:=\{l\}$, $REASON_l^i:=TRUE_i:=\{t\}$, $FALSE_i:=\{f\}$, $NECESSARY_i:=TRUE_i$, $<_i:=M_i\times M_i$. The symbols $+$ and $\cdot$ are again interpreted as the trivial, idempotent operations on $L_i$. The Gamma-function is defined simultaneously for all assignments $\gamma:V_P\cup V_J\rightarrow M_i\cup L_i$ as follows:

\begin{equation*}
\begin{split}
\varGamma_i(x,\gamma) &:=\gamma(x),\text{ for }x\in V_P\cup V_J\\
\varGamma_i(c)&:=l,\text{ for }c\in C\\
\varGamma_i(d)&:=t,\text{ for }d\in D\\
\varGamma_i(\varphi:true,\gamma)&:=\varGamma_i(\square\varphi,\gamma):=\varGamma_i(\varphi:s,\gamma):=\varGamma_i(\varphi,\gamma)\\
\varGamma_i(\varphi:false,\gamma)&:=\varGamma_i(\neg\varphi,\gamma):=
\begin{cases}
\begin{split}
&t,\text{ if }\varGamma_i(\varphi,\gamma)=f\\
&f,\text{ if }\varGamma_i(\varphi,\gamma)=t
\end{split}
\end{cases}
\end{split}
\end{equation*}

\begin{equation*}
\begin{split}
\varGamma_i(\varphi\rightarrow\psi,\gamma)&:=
\begin{cases}
\begin{split}
&t,\text{ if }\varGamma_i(\varphi,\gamma)=f\text{ or }\varGamma_i(\psi,\gamma)=t\\
&f,\text{ if }\varGamma_i(\varphi,\gamma)=t\text{ and }\varGamma_i(\varphi,\gamma)=f
\end{split}
\end{cases}
\end{split}
\end{equation*}

\begin{equation*}
\begin{split}
\varGamma_i(\varphi\equiv\psi,\gamma)&:=
\begin{cases}
\begin{split}
&t,\text{ if }\varGamma_i(\varphi,\gamma)=\varGamma_i(\psi,\gamma)\\
&f,\text{ else }
\end{split}
\end{cases}
\end{split}
\end{equation*}

\begin{equation*}
\varGamma_i(\varphi <\psi,\gamma):=\varGamma_i(s\le t,\gamma):=t
\end{equation*}

\begin{equation*}
\begin{split}
\varGamma_i(\bigwedge u.\varphi,\gamma):=
\begin{cases}
\begin{split}
&t,\text{ if }\varGamma_i(\varphi,\gamma_u^l)\in TRUE\\
&f,\text{ else }
\end{split}
\end{cases}
\end{split}
\end{equation*}

\begin{equation*}
\begin{split}
\varGamma_i(\forall x.\varphi,\gamma)&:=
\begin{cases}
\begin{split}
&t,\text{ if }\varGamma_i(\varphi,\gamma_x^m)\in TRUE\text{ for all }m\in M\\
&f,\text{ else }
\end{split}
\end{cases}
\end{split}
\end{equation*}

Similarly as in the proof of Theorem \ref{800} one shows that $\mathcal{M}_i$ satisfies all properties of a model. We now consider the specific assignments $\varrho_i:V_P\cup V_J\rightarrow M_i\cup L_i$ defined as follows:

\begin{equation*}
\begin{split}
\varrho_i(x)&:=
\begin{cases}
\begin{split}
&t,\text{ if }x\in T_i\text{ for }x\in V_P\\
&f,\text{ if }x\notin T_i\text{ for }x\in V_P\\
&l,\text{ if }x\in V_J
\end{split}
\end{cases}
\end{split}
\end{equation*}

We put $T'_i:=Th((\mathcal{M}_i,\varrho_i)):=\{\varphi\in Fm(C,D)\mid(\mathcal{M}_i,\varrho_i)\vDash\varphi\}$. Note that the $T'_i$ have the following property: $\varphi\in T'_i\Leftrightarrow\square\varphi\in T'_i$. This will be crucial for our construction.\\
\textbf{Claim} 2: $\mathcal{L}\cap F_N=\mathcal{L}\cap\bigcap_{i\in I} T'_i$.\\
\textit{Proof of the Claim}: By induction on $\varphi\in\mathcal{L}$ one shows: $\mathcal{L}\cap T_i\subseteq\mathcal{L}\cap T'_i$, for every $i\in I$. It follows that $\mathcal{L}\cap F_N=\mathcal{L}\cap\bigcap_{i\in I} T_i\subseteq\mathcal{L}\cap\bigcap_{i\in I} T'_i$. We now show $\mathcal{L}\cap\bigcap_{i\in I} T'_i\subseteq \mathcal{L}\cap\bigcap_{i\in I} T_i$ by induction on the formulas of $\mathcal{L}$. The only interesting case in the induction step is $\square\psi\in \mathcal{L}\cap\bigcap_{i\in I} T'_i$. Assuming this it follows that $\psi\in \mathcal{L}\cap\bigcap_{i\in I} T'_i$, and by induction hypothesis, $\psi\in \mathcal{L}\cap\bigcap_{i\in I} T_i$. Thus, $\psi\in\mathcal{L}\cap F_N$. That is, $\square\psi\in F_M$. Since the model $\mathcal{M}$ satisfies the truth condition (4), i.e. all formulas of the form $\square\varphi\rightarrow\square\square\varphi$, we get $\square\square\psi\in F_M$, that is, $\square\psi\in F_N$. Thus, $\square\psi\in\mathcal{L}\cap\bigcap_{i\in I} T_i$. This proves the Claim.\\

We define worlds $w_*:=\mathcal{L}\cap F_M$ and $w_i:=\mathcal{L}\cap T'_i$, for $i\in I$ (we suppose that $*\notin I$). The frame $(W,R)$ is given by $W:=\{w_*\}\cup\{w_i\mid i\in I\}$ and $R:=\{(w_*,w_i)\mid i\in I\}\cup\{(w_*,w_*)\}\cup\{(w_i,w_i)\mid i\in I\}$. The truth value assignment $g:W\rightarrow (V_P\rightarrow\{0,1\})$ is defined by $g_w(x)=1 :\Leftrightarrow x\in w$, for $w\in W$. \\
\textbf{Claim} 3: For all $\varphi\in\mathcal{L}$ and for all $w\in W$: $(w,g)\Vdash\varphi\Leftrightarrow\varphi\in w$.\\
\textit{Proof of the Claim}: The Claim can be proved by induction on $\varphi\in\mathcal{L}$, simultaneously for all $w_i$, $i\in I$, and then for $w_*\in W$. We show only the case $\varphi=\square\psi$. 
Let $i\in I$. Then
\begin{equation*}
\begin{split}
(w_i,g)\Vdash\square\psi &\Leftrightarrow (w_i,g)\Vdash\psi\\
&\Leftrightarrow\psi\in w_i,\text{ by induction hypothesis}\\
&\Leftrightarrow\psi\in\mathcal{L}\cap T'_i\\
&\Leftrightarrow\square\psi\in\mathcal{L}\cap T'_i\\
&\Leftrightarrow\square\psi\in w_i
\end{split}
\end{equation*} 

\begin{equation*}
\begin{split}
(w_*,g)\Vdash\square\psi &\Leftrightarrow (w_*,g)\Vdash\psi\text{ and }(w_i,g)\Vdash\psi\text{ for all }i\in I\\
&\Leftrightarrow\psi\in w_*\text{ and }\psi\in w_i\text{ for all }i\in I,\text{ by induction hypothesis }\\
&\Leftrightarrow\psi\in\mathcal{L}\cap F_M\text{ and }\psi\in\mathcal{L}\cap\bigcap_{i\in I} T'_i\\
&\Leftrightarrow\psi\in\mathcal{L}\cap F_N,\text{ by Claim 2}\\
&\Leftrightarrow\square\psi\in\mathcal{L}\cap F_M=w_*.
\end{split}
\end{equation*}
It is clear that $(W,R)$ is a frame of modal logic S4. It remains to prove the following \textbf{Claim} 4: For all $\varphi\in\mathcal{L}$: 
\begin{equation*}
(\mathcal{M},\beta)\vDash\varphi\Leftrightarrow (w_*,g)\Vdash\varphi.
\end{equation*}
\textit{Proof of the Claim}: This is again an induction on $\varphi\in\mathcal{L}$. We show the case $\varphi=\square\psi$:\\
$(\mathcal{M},\beta)\vDash\square\psi\Leftrightarrow\varGamma(\square\psi,\beta)\in TRUE\Leftrightarrow\square\psi\in F_M\Leftrightarrow (w_*,g)\Vdash\square\psi$, by Claim 3. Thus, Claim 4 is true. This finishes the proof of the theorem. $\square$

\end{document}